 \documentclass[final,5p,times,twocolumn]{elsarticle}

\usepackage{amssymb}
\usepackage{lipsum}

\usepackage{bm}
\usepackage{amsfonts}
\usepackage{latexsym}
\usepackage[utf8]{inputenc}
\usepackage{graphicx}
\usepackage{amsmath}
\usepackage{textcomp}
\usepackage{subfigure}
\usepackage{commath}
\usepackage{caption}
\usepackage{multirow}
\usepackage{ulem}
\usepackage{tabularx}
\usepackage{booktabs}  
\usepackage{xcolor}
\usepackage{hyperref}

\usepackage{float}
\usepackage{dcolumn}
\usepackage{ragged2e}
\usepackage{hyperref}
\usepackage{tensor}
\hypersetup{colorlinks,citecolor=blue}
\usepackage{aas_macros}
\usepackage{physics}

\journal{Journal of High Energy Astrophysics}

\begin{document}

\begin{frontmatter}



\title{Identifying highly magnetized white dwarfs: A dimensionality reduction framework for estimating magnetic fields}

\author[AOUW1]{Surajit Kalita\corref{cor1}}
\ead{skalita@astrouw.edu.pl}

\author[TDLI]{Akhil Uniyal}
\ead{akhil\_uniyal@sjtu.edu.cn}

\author[AOUW1]{Tomasz Bulik}
\ead{tb@astrouw.edu.pl}

\author[TDLI,addr2,addr3]{Yosuke Mizuno}
\ead{mizuno@sjtu.edu.cn}

\affiliation[AOUW1]{%
  organization={Astronomical Observatory, University of Warsaw}, 
  addressline={Al. Ujazdowskie 4},
  city={Warszawa},
  postcode={PL-00478},
  country={Poland}
}

\affiliation[TDLI]{%
  organization={Tsung-Dao Lee Institute, Shanghai Jiao Tong University},
  addressline={1 Lisuo Road},
  city={Shanghai},
  postcode={201210},
  country={People’s Republic of China}
}

\affiliation[addr2]{%
  organization={School of Physics and Astronomy, Shanghai Jiao Tong University},
  addressline={800 Dongchuan Road},
  city={Shanghai},
  postcode={200240},
  country={People’s Republic of China}
}

\affiliation[addr3]{%
  organization={Key Laboratory for Particle Astrophysics and Cosmology (MOE) and Shanghai Key Laboratory for Particle Physics and Cosmology},
  addressline={800 Dongchuan Road},
  city={Shanghai},
  postcode={200240},
  country={People’s Republic of China}
}

\begin{abstract}
Magnetic fields play a crucial role in compact object physics, particularly in white dwarfs (WDs), where high densities can sustain strong magnetic fields. Observations have revealed magnetized WDs (MWDs) with surface fields reaching approximately $10^9$\,G, although high-field MWDs are fewer in number in current catalogs owing to their intrinsic faintness and limitations in conventional electromagnetic surveys. In this study, we apply unsupervised machine learning (ML) techniques to systematically analyze a sample of hydrogen-atmosphere (DA) WDs. Using Uniform Manifold Approximation and Projection (UMAP) for dimensionality reduction and Density-Based Spatial Clustering of Applications with Noise (DBSCAN) for cluster identification, we classify distinct subpopulations within the DA WD sample. Each cluster exhibits unique intrinsic properties such as mass, surface gravity, temperature, and age. Our analysis further reveals that these subgroups effectively differentiate MWDs from non-magnetic or weakly magnetic counterparts. Moreover, utilizing a set of previously confirmed MWDs, we estimate the field strengths of all other MWDs lacking magnetic field measurements. This study underscores the effectiveness of ML-based approaches in astrophysical discovery, particularly detecting magnetized compact objects when direct measurements are unavailable.
\end{abstract}




\end{frontmatter}




\section{Introduction}
White dwarfs (WDs) are the final evolutionary stages of low- to intermediate-mass stars, typically originating from progenitors with initial masses $\lesssim 10\pm2 \rm\, M_\odot$~\citep{2018MNRAS.480.1547L}. Their gravitational pull is primarily supported by the electron degeneracy pressure. The seminal work of \cite{1935MNRAS..95..207C} established that this pressure balance results in an inverse mass--radius relation, leading to a maximum stable mass of about $1.4\,\rm M_\odot$ for non-rotating, non-magnetized configurations, famously known as the Chandrasekhar mass limit. While no WD has been directly observed to exceed this limit, recent detections of several over-luminous Type Ia supernovae may suggest the existence of super-Chandrasekhar progenitors~\citep{2006Natur.443..308H,2010ApJ...713.1073S}, possibly due to rapid rotation, strong magnetic fields, or other exotic physics~\citep{2020ApJ...896...69K,2021MNRAS.508..842K,2022PhRvD.105b4028S}.

Magnetic fields are a ubiquitous feature of compact objects, yet their detection in WDs remains observationally challenging. magnetized WDs (MWDs) are typically identified through signatures such as Zeeman-split spectral lines, circular polarization, and suppression of continuum flux in specific photometric bands~\citep{2007ApJ...654..499K,2012A&A...545A..30L,2015SSRv..191..111F,2023ApJ...944...56A}. The strongest measured surface magnetic field in a WD is approximately $10^9\,\rm G$~\citep{2015SSRv..191..111F,2000MNRAS.317..310H}, and theoretical considerations imply that the internal field strengths may be significantly stronger because of higher density~\citep{2014MNRAS.437..675W}. Such extremely high fields might have non-negligible consequences on the structural and thermal properties of WDs. In general, MWDs are typically more massive than their non-magnetic counterparts as magnetic pressure provides additional support against gravity~\citep{2019MNRAS.490.2692K}. Moreover, MWDs are relatively less luminous because strong fields are believed to obstruct convective energy transport within the stellar envelope~\citep{2024MNRAS.529.4577K}. Magnetic fields also modify the equation of state, shape, and size of the WD, and impact the cooling rates~\citep{2013PhRvL.110g1102D,2024MNRAS.534L..65G}. Despite their astrophysical significance, only a small fraction of WDs have been confirmed as highly magnetized~\citep{2017ASPC..509....3D}, likely due to observational biases favoring brighter, less magnetized objects.

The origin of strong magnetic fields in WDs remains largely unknown. The most widely accepted theory is flux freezing, where fields from main-sequence progenitors are amplified during stellar contraction~\citep{2015SSRv..191..111F,1981ApJS...45..457A}. An alternative scenario states that binary mergers may generate MWDs with fields via differential rotation and dynamo amplification~\citep{2008MNRAS.387..897T,2012ApJ...749...25G,2015MNRAS.447.1713B}. This is also supported by observational evidence of higher average mass of MWDs~\citep{2013MNRAS.429.2934K}, highlighting their origin via binary mergers. A systematic understanding of WD magnetism is thus necessary not only for WD astrophysics but also for stellar evolution and compact binary dynamics. Nevertheless, measuring magnetic fields in WDs is observationally challenging, and as a result, such measurements exist for only a small fraction of the entire population. It reflects on the current observational datasets on magnetic field strength of WDs, which are largely incomplete, necessitating robust data-driven models approach to infer field strengths from other observables.

Large-volume astronomical surveys provide vast datasets but often suffer from the `curse of dimensionality'~\citep{2014sdmm.book.....I}, a phenomenon commonly referred to when diverse sampling in high-dimensional parameter spaces obscures underlying physical trends. However, these drawbacks are mitigated in low-dimensional space. A commonly adopted approach to address these issues is dimensionality reduction, which aims to identify intrinsic patterns embedded in high-dimensional datasets. These low-dimensional representations enable easy visualization and improved intelligibility of the physical patterns. Over the past decades, a range of machine learning (ML) algorithms have been developed to address the challenge of high-dimensional data representation, such as Principal Component Analysis (PCA), t-Distributed Stochastic Neighbor Embedding (t-SNE), Uniform Manifold Approximation and Projection (UMAP), and many more, which help in recognizing meaningful trends in complex datasets~\citep{2006Sci...313..504H}. Recent works on WDs have demonstrated that the unsupervised ML methods such as UMAP and self-organizing maps can be really useful for the detection of metal-polluted systems or systems with unresolved companions~\citep{2024ApJ...970..181K,2024ApJ...977...31P,2025ApJ...988...51P,2025A&A...701A.273L}. On the other hand, the combination of supervised algorithms like Random Forests and neural posterior estimation methods enables not only robust spectral/photometric classification but also rapid and fully probabilistic inference of physical parameters over the large-survey catalogs~\citep{2024MNRAS.535.2246B,2025A&A...699A...3G,2025MNRAS.544.1939V}. In other branches of astronomy, these techniques have been used to classify fast radio bursts~\citep{2025ApJ...982...16Q,2026JHEAp..4900449J}, constrain cosmological parameters~\citep{2021A&A...648A..74H}, classify kinematic morphologies of galaxies~\citep{2023A&A...671A..19R}, black-hole imaging techniques~\citep{2022PhRvD.106b3017C}, classify gamma-ray sources~\citep{2016ApJ...820....8S}, and the list goes on.

As the number of WDs with measured magnetic field strengths is small, this study proposes a comprehensive data-driven ML framework to estimate surface magnetic field strengths from other intrinsic stellar parameters to complement the current MWD catalogs. We apply the UMAP technique~\citep{2018arXiv180203426M} for dimensionality reduction of the input parameter space and Density-Based Spatial Clustering of Applications with Noise (DBSCAN) algorithm~\citep{1996kddm.conf..226E} for identifying subpopulations of hydrogen-atmosphere (DA) WDs with similar physical characteristics. Within the resulting clusters that contain known MWDs, we perform k-Nearest Neighbors (kNN) regression to infer magnetic field strengths for the remaining magnetized sources lacking direct field measurements.

The remainder of this paper is organized as follows. In Section~\ref{Sec2}, we present a detailed description of the WD dataset used in our study, including the methodology for dimensionality reduction and the applied clustering framework. Section~\ref{Sec3} outlines the results, provides a comprehensive analysis of the identified clusters, and examines the influence of the intrinsic stellar properties on the clustering structure. We further investigate the distribution of magnetic field strengths within these clusters in Section~\ref{Sec4}, allowing us to estimate the magnetic field strength of MWDs without direct measurements. We explain the technical details of each ML technique used in this study in~\ref{Appendix}. Finally, Section~\ref{Sec5} summarizes our key findings and discusses their broader implications for future studies on stellar magnetism and WD population analysis with concluding remarks are put in Section~\ref{Sec6}.

\section{Dataset description and methodological framework}\label{Sec2}

In this study, we consider DA WDs cataloged in the Montreal White Dwarf Database (MWDD)\footnote{\url{https://www.montrealwhitedwarfdatabase.org/home.html}}. We select six fundamental stellar parameters to investigate the intrinsic properties of WDs, as listed in Table~\ref{Tab: Parameters}. Although the MWDD catalog contains 20\,498 DA WDs, only 1\,043 have complete entries for all six parameters\footnote{As of July 2025}. To ensure consistency, we restrict our study to this subset, maintaining a homogeneous dataset for robust statistical analysis.
\begin{table}[htpb]
    \centering
    \caption{Intrinsic parameters of WDs reported in MWDD catalog.}
    \begin{tabular}{|l|l|}
        \hline
        Parameter & Description \\
        \hline
        $M$ & Mass of WD \\
        $\log g$ & Surface gravity of WD in logarithmic scale \\
        $T_\mathrm{eff}$ & Effective surface temperature \\
        $\log L$ & Luminosity in logarithmic scale \\
        $\tau$ & Cooling age of WD \\
        $\mathsf{M}_V$ & Absolute magnitude \\
        \hline
    \end{tabular}
    \label{Tab: Parameters}
\end{table}

Among the six basic parameters characterizing WDs, the surface gravity ($\log g$) and effective temperature ($T_\mathrm{eff}$) are typically determined from the spectroscopic broadening of hydrogen Balmer lines using synthetic model atmospheres of DA WDs~\citep{1992ApJ...394..228B,2009ApJ...696.1755T,2011ApJ...743..138G}. Alternatively, $T_\mathrm{eff}$ can be estimated photometrically by fitting the observed spectral energy distribution (SED) to theoretical models across ultraviolet to infrared wavelengths~\citep{2006AJ....132.1221H,2019ApJ...871..169G}. In this context, $T_\mathrm{eff}$ refers to the temperature of a blackbody emitting the same total flux as that of the observed WD.

The mass of a WD ($M$) is not directly measurable; rather, it is inferred by combining spectroscopically determined values of $\log g$ and $T_\mathrm{eff}$ with theoretical mass--radius relations derived from the stellar evolutionary models~\citep{2001PASP..113..409F,2010A&ARv..18..471A,2017MNRAS.465.2849T}. Similarly, bolometric luminosity $L$ can be obtained using the Stefan-Boltzmann law. Alternatively, for WDs with precise distance measurements, such as in the case of $Gaia$ parallaxes, $L$ can be derived from the absolute bolometric magnitude~\citep{2019MNRAS.482.4570G}.

The cooling age of a WD ($\tau$) is inferred using its $T_\mathrm{eff}$ and $M$ on theoretical cooling tracks with different mechanisms such as radiative cooling, neutrino emission, and crystallization, along with core composition assumptions~\citep{1995LNP...443...41W,2001PASP..113..409F,2010A&ARv..18..471A}. Estimating the total stellar age further requires extrapolation to the initial mass of the progenitor, which is usually obtained via empirical initial--final mass relations~\citep{2008ApJ...676..594K,2018ApJ...866...21C}. In the MWDD catalog, the reported cooling age represents the duration since it entered the WD phase. Furthermore, its absolute visual magnitude ($\mathsf{M}_V$) is determined from apparent magnitude and distance modulus. In cases where a direct distance measurement is unavailable, the theoretical distance and corresponding luminosity can be estimated from $\log g$ and $T_\mathrm{eff}$. 

\begin{figure}[htpb]
    \centering
    \includegraphics[scale=0.5]{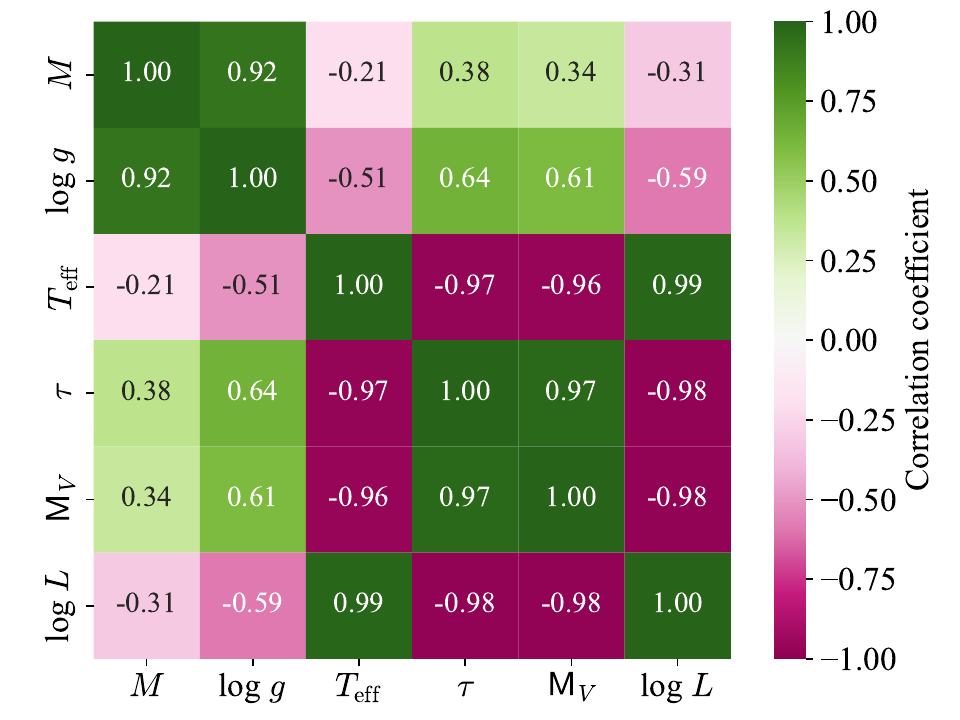}
    \caption{Spearman correlation matrix for different parameters of WDs mentioned in Table~\ref{Tab: Parameters}.}
    \label{Fig: Spearman}
\end{figure}

Note that many of these parameters are interconnected through different theoretical models. For instance, $\mathsf{M}_V$ can be related to $M$ and $L$ of the WD. However, if the V-band magnitude is not measured for a WD, $\mathsf{M}_V$ is not reported. Conversely, when $\mathsf{M}_V$ is available, the absence of a bolometric correction or detailed spectral fit may prevent the determination of $L$ and, consequently, its cooling age. A Spearman corelation matrix among these parameters is shown in Fig.~\ref{Fig: Spearman}. To ensure consistency of parameters, this analysis is strictly restricted to WDs for which all six parameters are available, irrespective of the parameters being correlated or uncorrelated.

\begin{figure}[htpb]
    \centering
    \includegraphics[scale=0.5]{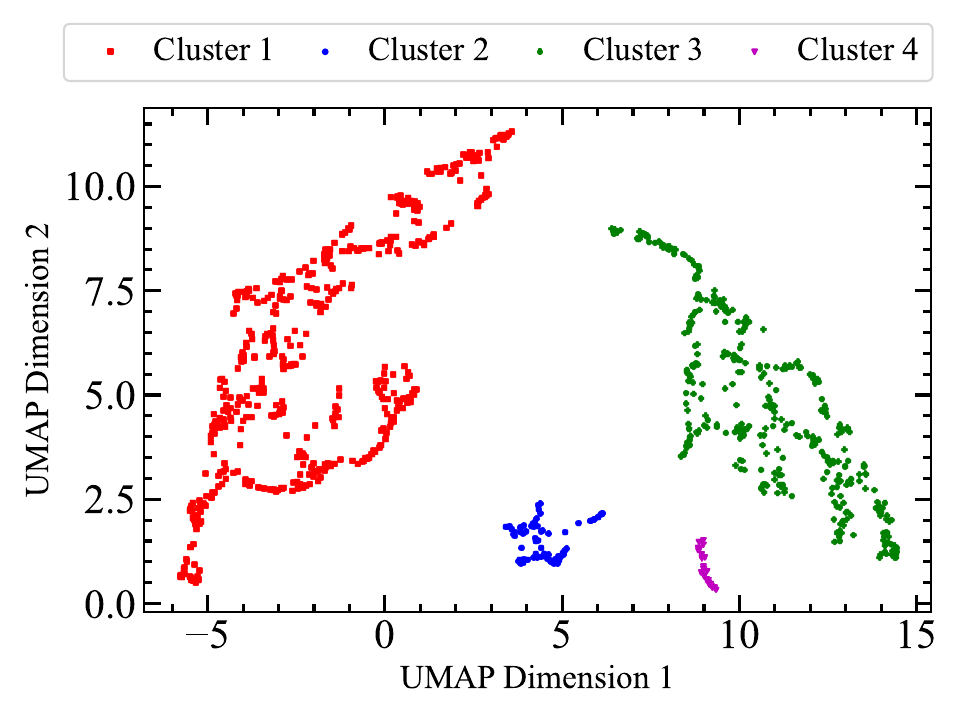}
    \caption{2-dimensional projection of the WD dataset using the UMAP algorithm and subsequent clustering using the DBSCAN algorithm. The visualization reveals four distinct clusters. Each point represents a WD, colored according to its assigned cluster.}
    \label{Fig: UMAP+DBScan}
\end{figure}

\begin{figure*}[htpb]
	\centering
	\subfigure[Mass]{\includegraphics[scale=0.5]{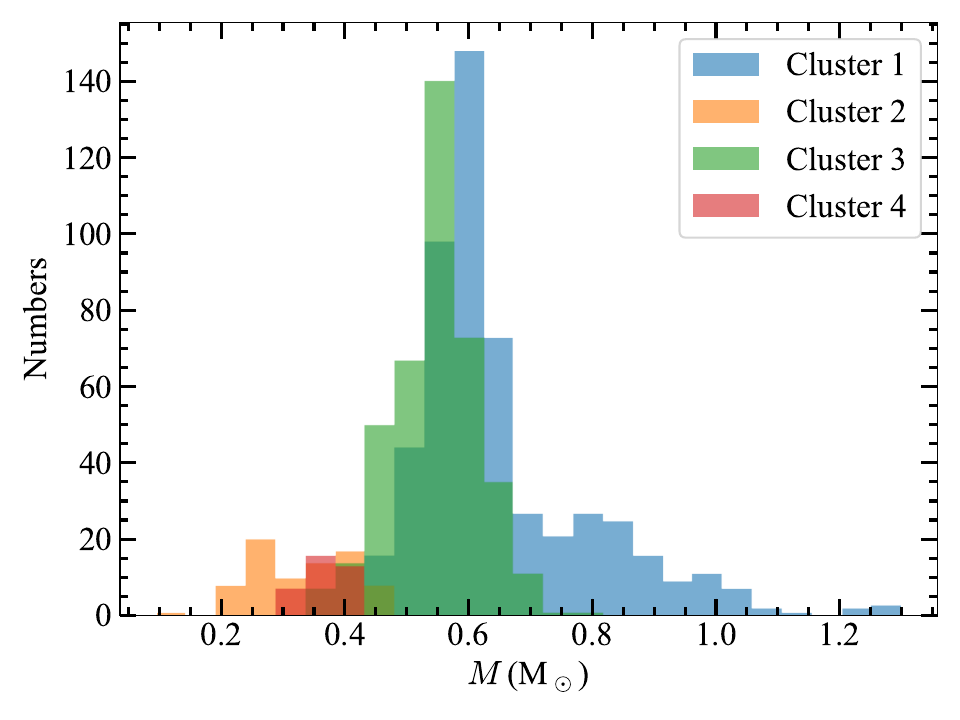}}
	\subfigure[Surface gravity]{\includegraphics[scale=0.5]{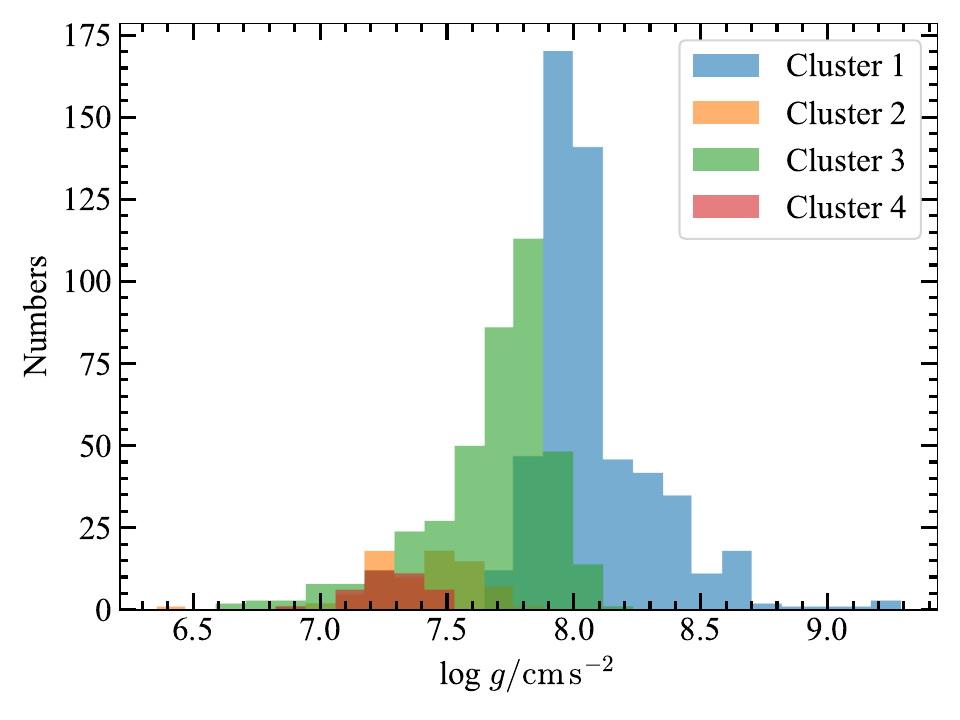}}
	\subfigure[Effective temperature]{\includegraphics[scale=0.5]{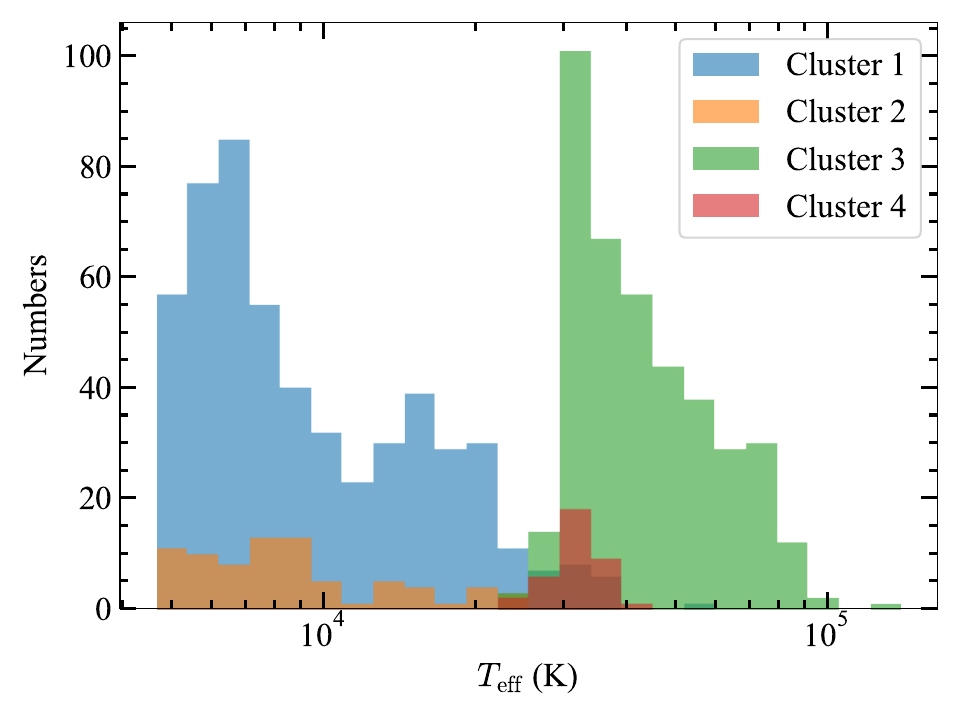}}
	\subfigure[Luminosity]{\includegraphics[scale=0.5]{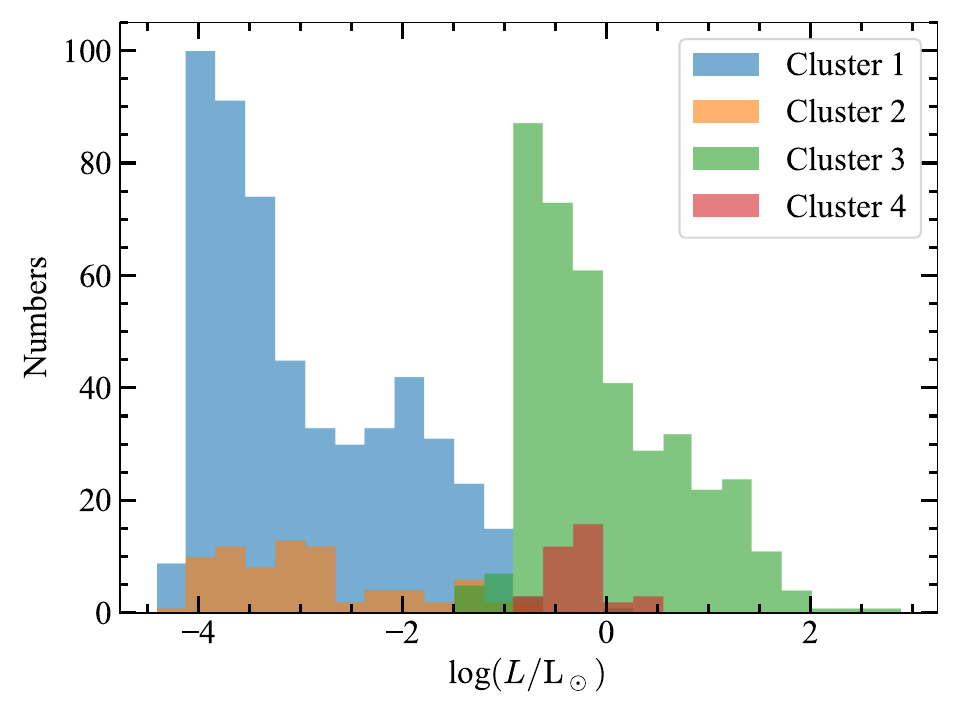}}
	\subfigure[Cooling age]{\includegraphics[scale=0.5]{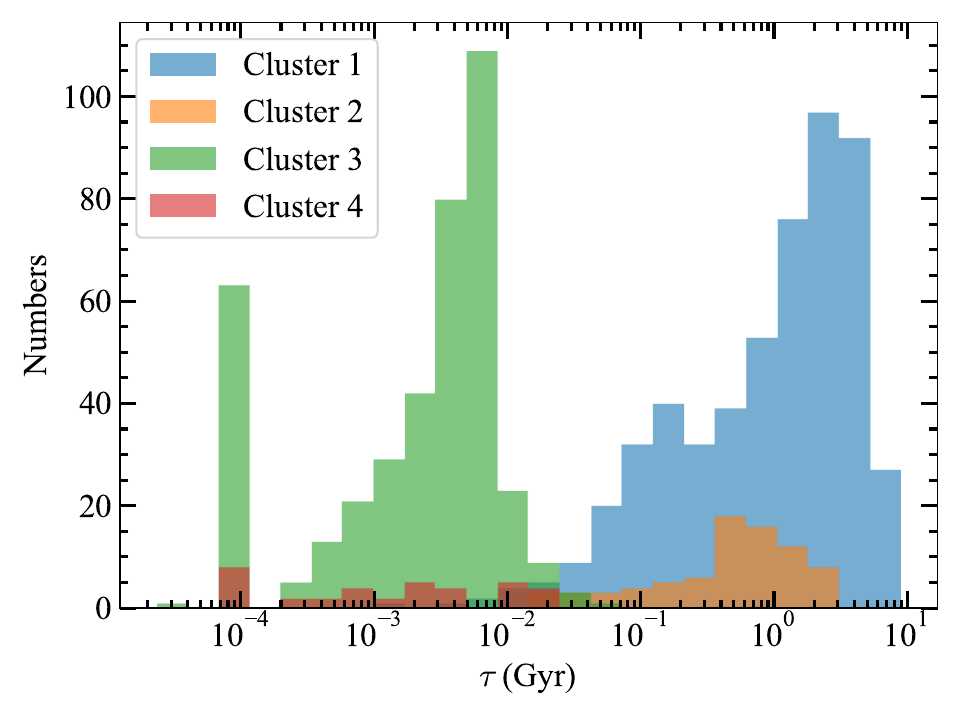}}
	\subfigure[Absolute magnitude]{\includegraphics[scale=0.5]{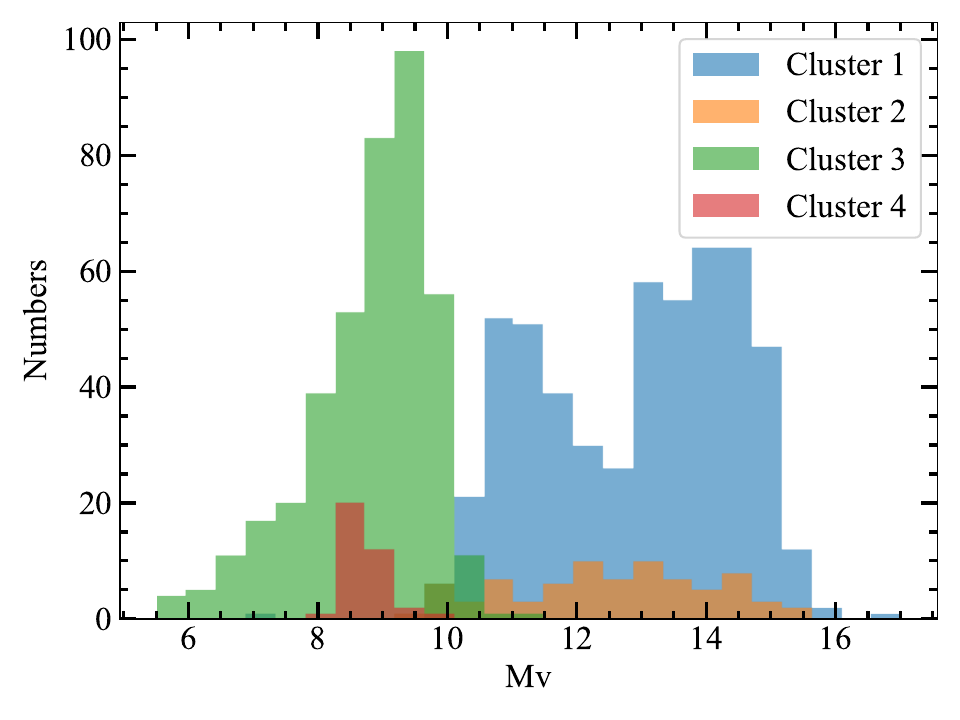}}
	\caption{Histograms showing intrinsic properties of WDs across different clusters, based on data of Fig.~\ref{Fig: UMAP+DBScan}.}
	\label{Fig: Histograms}
\end{figure*}

As visualizing a 6-dimensional parameter space is not feasible, we transpose the problem into lower dimensions to allow easier analysis. In this study, we adopt the UMAP algorithm~\citep{2018arXiv180203426M}, a non-linear manifold learning technique that preserves both local and global structures during dimensionality reduction. UMAP projects the high-dimensional data onto a lower-dimensional manifold, retaining topological features essential for identifying physical correlations and clustering behavior. UMAP is a powerful dimensionality-reduction method because it is not required to mention beforehand whether the input features are correlated or uncorrelated, as described in~\ref{Appendix}. In other words, the selected six parameters are not required to be statistically independent; any correlations or functional dependencies among them do not affect the outcome of the analysis, as the method is inherently robust to such parameter interdependence. Fig.~\ref{Fig: UMAP+DBScan} displays the projection of the original 6-dimensional dataset over a 2-dimensional manifold using the UMAP algorithm. We further use the DBSCAN algorithm in the projected space, revealing four well-separated, internally coherent clusters, implying a robust internal grouping within the sample. Clusters 1, 2, 3, and 4 consist of 530, 78, 399, and 36 data points, respectively. Table~\ref{Tab: Values} lists the cluster-wise details of all six original parameters, and Fig.~\ref{Fig: Histograms} shows the histogram distributions of these properties. The differences in these parameters among the clusters are considerable, indicating that all six features play significant roles in cluster separation. This highlights the multidimensional nature of data structuring and the relevance of each parameter in forming distinct sub-populations within the dataset.

\begin{table*}[htpb]
    \centering
    \caption{Mean values of intrinsic parameter within each cluster.}
    \begin{tabular}{|l|l|l|l|l|l|l|}
        \hline
        Cluster & $M$ & $\log g/\rm cm\,s^{-2}$ & $T_\mathrm{eff}$ & $\log L/\rm L_\odot$ & $\tau$ & $\mathsf{M}_V$\\
        & ($\mathrm{M}_\odot$) & & ($10^4$\,K) & & (Myr) & \\
        \hline
        1 & 0.66 & 8.09 & 1.09 & -2.989 & 1762.99 & 12.99\\
        2 & 0.33 & 7.40 & 0.95 & -2.885 & 769.24 & 12.52\\
        3 & 0.55 & 7.67 & 4.57 & -0.001 & 4.39 & 8.82\\
        4 & 0.37 & 7.28 & 3.16 & -0.248 & 4.49 & 8.70\\
        \hline
    \end{tabular}
    \label{Tab: Values}
\end{table*}

\section{Cluster analysis and stellar property correlations}\label{Sec3}

\begin{figure*}[htpb]
	\centering
	\subfigure[Mass]{\includegraphics[scale=0.5]{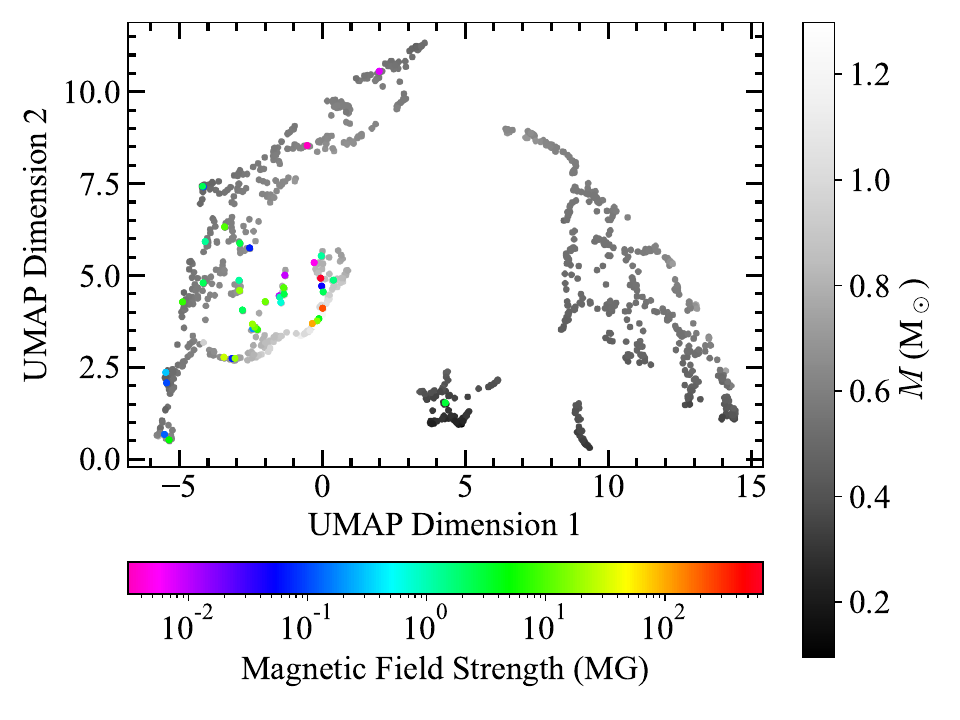}}
	\subfigure[Surface gravity]{\includegraphics[scale=0.5]{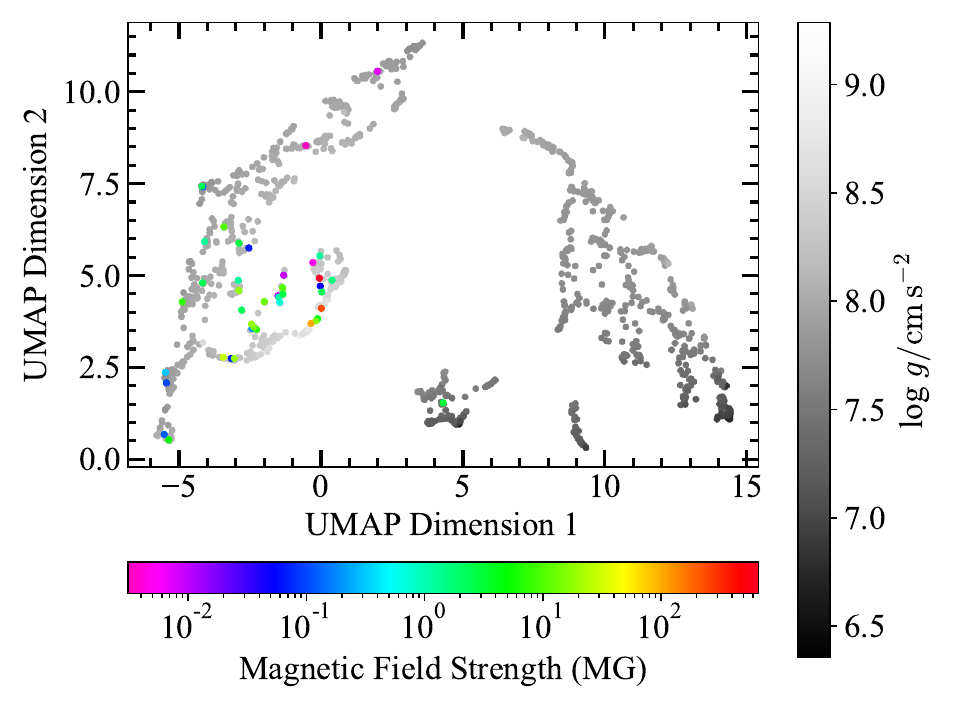}}
	\subfigure[Effective temperature]{\includegraphics[scale=0.5]{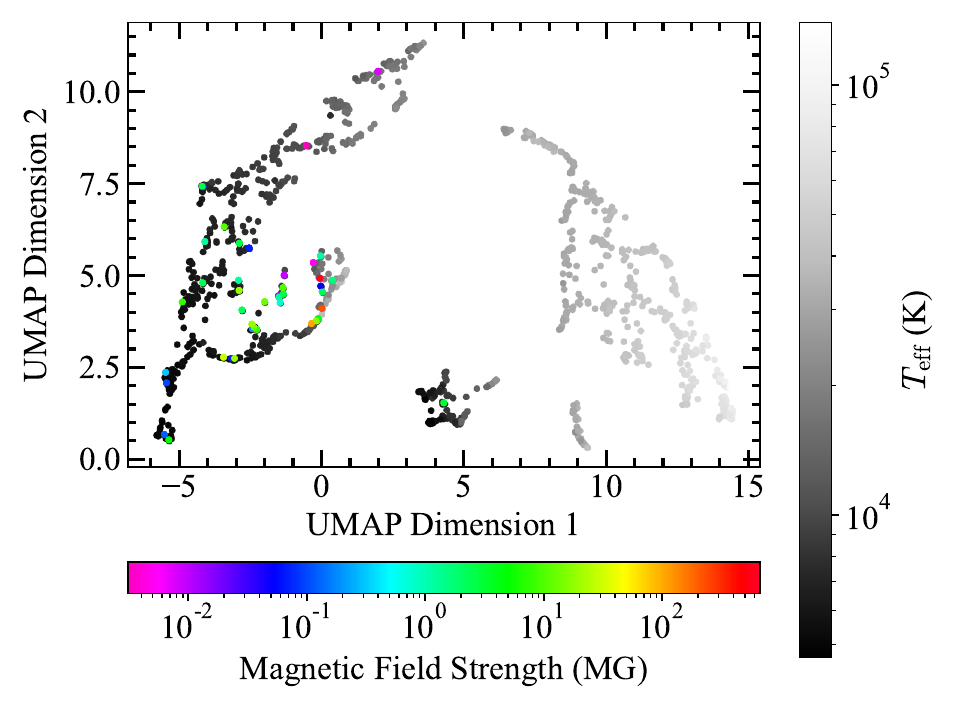}}
	\subfigure[Cooling age]{\includegraphics[scale=0.5]{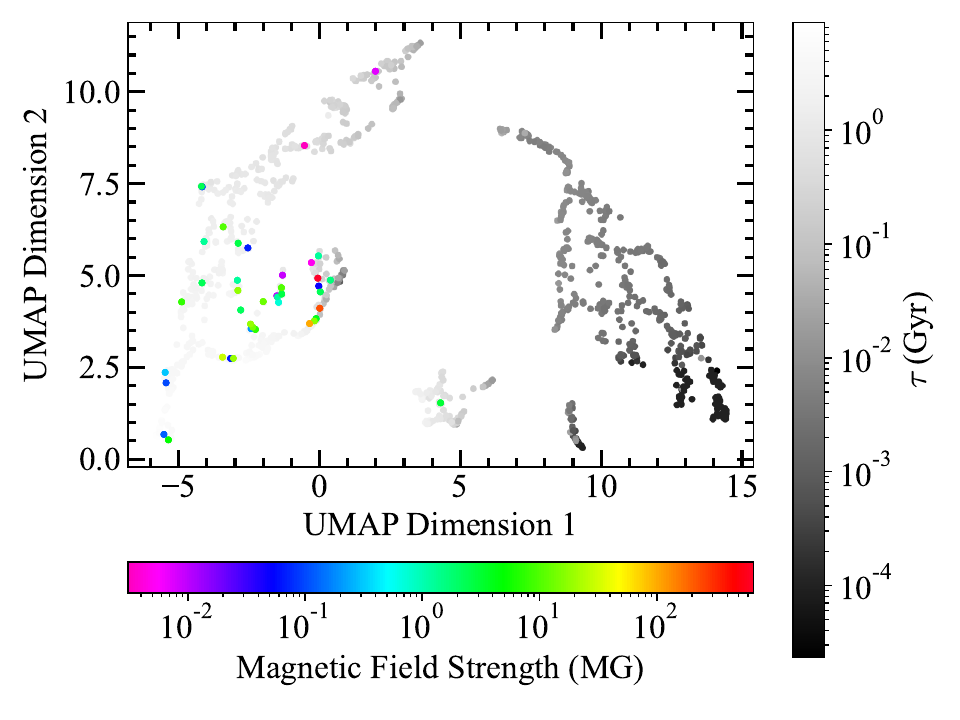}}
	\caption{UMAP projection of WD dataset annotated with confirmed MWDs and their surface magnetic field strengths. The vertical color bar in each subplot shows the values of the corresponding parameter, while the horizontal color bar shows the magnetic field strengths. The magnetized WDs are predominantly concentrated within Cluster 1.}
	\label{Fig: All parameters}
\end{figure*}

We now examine the distribution of WDs with reported magnetic fields (as listed in Table~\ref{Tab: Confirmed magnetic}) within the dataset, plotting their surface magnetic field strengths $B$ on the 2-dimensional UMAP projected space, as shown in Fig.~\ref{Fig: All parameters}. This reveals an interesting pattern: nearly all confirmed MWDs are grouped within Cluster 1, except for an outlier in Cluster 2. Clusters 3 and 4 contain no known MWDs, implying that the magnetic fields and six underlying intrinsic parameters are closely related\footnote{Note that we do not claim the WDs in clusters 3 and 4 are non-magnetic. Given that magnetic field is ubiquitous in astronomical objects, our result instead indicates these WDs, on average, might possess weaker field strengths than those in cluster 1.}. Moreover, within Cluster 1, MWDs mostly occupy the bottom of this cluster, and those with strong magnetic fields tend to be distributed along a particular branch of this cluster. This structured spatial segregation suggests that the magnetic field strength depends on the combined influence of the six intrinsic stellar parameters selected in the present analysis. Such complex multidimensional correlations would remain undistinguished in traditional pairwise parameter plots, demonstrating the unique capability of non-linear dimensionality reduction techniques like UMAP for uncovering hidden relationships in high-dimensional astrophysical data.

\begin{table}[htpb]
\centering
\caption{MWDs with reported magnetic fields.}
\label{Tab: Confirmed magnetic}
\begin{tabular}{|l|l|}
\hline
WD name & Reported $B$ (MG) \\
\hline
WD\,J$232847.64+051454.24$ & 0.0031 \\
WD\,J$164825.63+590322.66$ & 0.005 \\
WD\,J$041521.80-073929.20$ & 0.0073 \\
WD\,J$211316.85-814912.88$ & 0.009 \\
WD\,J$000210.72-430955.39$ & 0.0098 \\
WD\,J$204421.46-680521.36$ & 0.05 \\
WD\,J$195629.23-010232.67$ & 0.07 \\
WD\,J$165445.69+382936.63$ & 0.07 \\
WD\,J$182119.83+610107.38$ & 0.1 \\
WD\,J$025959.15+081156.43$ & 0.1 \\
WD\,J$032511.05-014915.05$ & 0.12 \\
WD\,J$001214.75+502520.74$ & 0.2 \\
WD\,J$073330.88+640927.44$ & 0.3 \\
WD\,J$024208.44+111233.00$ & 0.725 \\
WD\,J$053714.90+675950.51$ & 1.1340 \\
WD\,J$064112.82+474419.65$ & 1.2 \\
WD\,J$065123.03+624254.29$ & 1.2180 \\
WD\,J$222740.36+175321.36$ & 1.3003 \\
WD\,J$122314.52+230751.04$ & 1.6073 \\
WD\,J$051553.54+283916.81$ & 2.2191 \\
WD\,J$150549.31-071440.95$ & 2.3 \\
WD\,J$060158.71+372601.73$ & 2.3001 \\
WD\,J$171450.80+391837.43$ & 2.3956 \\
WD\,J$233203.52+265846.12$ & 2.5684 \\
WD\,J$103532.60+212604.53$ & 2.6282 \\
WD\,J$182030.93+123921.0$ & 3 \\
WD\,J$165948.42+440104.04$ & 3.9026 \\
WD\,J$130841.20+850228.16$ & 4.9 \\
WD\,J$133250.72+011706.32$ & 7.36 \\
WD\,J$050552.46-172243.48$ & 7.4173 \\
WD\,J$134724.37+102138.01$ & 10 \\
WD\,J$135315.53-091633.37$ & 10 \\
WD\,J$164057.15+534109.32$ & 13 \\
WD\,J$181608.87+245442.85$ & 14 \\
WD\,J$225726.12+075542.60$ & 16.17 \\
WD\,J$064926.55+752124.97$ & 18.2428 \\
WD\,J$001412.79-131101.09$ & 18.7663 \\
WD\,J$102907.46+112719.28$ & 19.1813 \\
WD\,J$055625.46+052148.44$ & 28.9870 \\
WD\,J$153634.92-055240.02$ & 31 \\
WD\,J$190010.25+703951.42$ & 100 \\
WD\,J$103349.20+230916.26$ & 230.4932 \\
WD\,J$094846.64+242125.88$ & 670 \\
\hline
\end{tabular}
\end{table}

\subsection{Magnetic field strength distributions and identification of highly magnetized white dwarfs}\label{Sec4}


We now restrict ourselves within Cluster 1 and 2 to estimate the magnetic field strengths of MWDs for which direct measurements are currently unavailable. Using confirmed MWDs as our training set, we apply kNN regression in the UMAP-reduced dimensionality space to estimate the zeroth-order magnetic field strengths of the remaining WDs. As UMAP preserves local and global structures of the data, it allows the kNN algorithm to effectively extrapolate the local continuity and morphological similarity among WDs. 

kNN regression method is effective in capturing meaningful local trends in the parameter space, even in the absence of direct magnetic field measurements. This implies that the magnetic field strength can be statistically modeled as a function of the six intrinsic parameters. The estimated field values of MWDs without direct field measurements are provided in Table~\ref{Tab: Magnetic}. As the catalog does not specify the uncertainties for the magnetic field values, we assume a 10\% uncertainty for each known magnetic field of the MWDs.
Our analysis reveals one MWD, WD\,J$023619.57+524412.41$, to possess a field strength of more than $10^8\rm\,G$, which has not yet been possible using conventional electromagnetic observations alone. This demonstrates the power of data-driven approaches to complement traditional observational methods in uncovering extreme astrophysical objects.

\begin{table*}[htpb]
    \centering
    \caption{Estimated magnetic fields of confirmed MWDs without fields measurements.}
    \begin{tabular}{|l|l|l|}
        \hline
        WD name & Spectral type & Estimated $B$ (MG) \\
        \hline
        WD\,J$023619.57+524412.41$ & DAH & $143.10\pm264.02$ \\
        WD\,J$001349.89-714954.26$ & DAH & $2.85\pm4.02$ \\
        WD\,J$105544.96+211105.58$ & DAH & $9.00\pm7.82$ \\
        WD\,J$092213.50+050438.31$ & DAH & $16.99\pm11.93$ \\
        WD\,J$155857.70+041705.16$ & DAH & $9.54\pm10.96$ \\
        WD\,J$151625.10+280320.94$ & DAH & $4.43\pm5.61$ \\
        WD\,J$110747.90-342051.49$ & DAH & $44.80\pm166.66$ \\
        WD\,J$122619.77+183634.46$ & DAHe & $3.17\pm4.33$ \\
        WD\,J$131925.56-782328.17$ & DAH & $18.36\pm11.89$ \\
        WD\,J$012403.99-424038.43$ & DAH & $61.72\pm160.15$ \\
        WD\,J$023521.79-240047.12$ & DAH & $66.17\pm165.02$ \\
        WD\,J$204906.70+372814.05$ & DAP & $68.75\pm194.03$ \\
        WD\,J$000728.90+340339.69$ & DAP & $10.92\pm9.45$ \\
       \hline
    \end{tabular}
    \label{Tab: Magnetic}
\end{table*}

In principle, this analysis could be extended to the entire sample of WDs to determine their magnetic field strengths. However, as evident from Fig.~\ref{Fig: All parameters}, the magnetic fields of confirmed MWDs exhibit a scattered distribution in the UMAP projection, indicating a degree of randomness in their embedding positions. 
Thus, we consider it inappropriate to assign magnetic field values to all WDs, as such extrapolation may yield unreliable outcomes. A more rigorous approach would be to take the spectroscopic data for each WD and combine it with other physical parameters to obtain more precise field estimates. A detailed spectrum-informed analysis will be conducted in future studies.

\section{Discussion}\label{Sec5}

This study establishes a comprehensive data-driven ML framework for estimating WD magnetic field strengths through correlated observable parameters. By integrating dimensionality reduction with UMAP, clustering with DBSCAN, and regression with kNN, we have demonstrated a robust approach to infer magnetic fields in the absence of direct measurements. By analyzing the intrinsic features of these WDs, we have identified and categorized subsets of the DA WD population obtained from the MWDD catalog. With the help of previously confirmed MWDs, we have assigned magnetic field strengths to all other MWDs without magnetic field measurements using the kNN regression method and have provided an empirical formula to estimate field values from their intrinsic stellar features: mass, surface gravity, luminosity, cooling age, surface temperature, and absolute magnitude. In this way, we have identified WD\,J$023619.57+524412.41$ that might possess strong magnetic fields with $B>10^8\rm\,G$. Examining its spectra, it resembles WD\,J$094846.64+242125.88$, whose magnetic field was measured to be 670\,MG directly from observations, which we have used in our training dataset. While the kNN method provides valuable first-order estimates, we note significant uncertainties due to local interpolation methods. In fact, kNN performs interpolation by considering only the spatial coordinates of the features in UMAP-reduced space. Hence, when the measured data become sparse in certain areas and they have non-uniform sampling density, this method increases the uncertainties. Additionally, if the underlying relationship between the input features and magnetic field is inherently non-linear, a simple distance-weighted kNN approach might not capture all such complexities, resulting in a high variance in the predictions. Thus, although this methodology may not capture all the complexities behind magnetic field generation and evolution in WDs, it certainly provides a robust starting point for empirically classifying MWDs based only on their intrinsic physical parameters.

Nevertheless, these ML methodologies independently recover established physical properties of MWDs, further validating the effectiveness of the dimensionality reduction and clustering approach. Fig.~\ref{Fig: All parameters} illustrates that highly magnetic WDs preferentially appear in regions with high mass and high surface gravity, consistent with their compact size and enhanced density~\citep{2019MNRAS.490.2692K,2023ApJ...944...56A,2023MNRAS.520.6111H}. This high density favors the generation and retention of strong magnetic fields, either by the process of fossil-field preservation~\citep{1981ApJS...45..457A} or by potential amplification mechanisms during post-main-sequence evolution~\citep{2014MNRAS.437..675W}. In addition, this analysis confirms that high-field WDs exhibit effectively lower temperatures and luminosities. This agrees with previous results, which suggest that magnetic WDs undergo fast cooling due to magnetic suppression of convective transport~\citep{2015SSRv..191..111F,2015ApJ...812...19T,2024MNRAS.529.4577K}, making these objects intrinsically fainter. Moreover, this analysis reveals that the strongest fields appear in older WDs. Although the physical mechanism is unknown, it has been proposed that WDs become more magnetized as they cool and grow older~\citep{2015ApJ...812...19T}. This may be related to the fact that high electrical conductivity significantly suppresses magnetic field decay~\citep{2002MNRAS.333..589C}. An alternative explanation involves the structural evolution of the WD core as it crystallizes over time, and internal restructuring might contribute to further amplification of existing magnetic fields~\citep{2017ASPC..509..409I,2024A&A...691L..21C}. These results highlight the observational bias against detecting high-field MWDs due to their intrinsic faintness which makes them challenging targets for conventional EM surveys. Our ML approach effectively compensates for this selection effect, revealing a hidden population of MWDs.

\section{Conclusions}\label{Sec6}

This study essentially demonstrates that a hierarchical ML data-driven approach, combining global dimensionality reduction with local regression techniques, provides an effective framework for predicting magnetic field strengths across WD populations. Our methodology offers a robust alternative in scenarios when direct magnetic field measurements are observationally prohibitive. The predicted high-field MWD candidates present prime targets for follow-up in future observations, enabling a more efficient allocation of telescope resources. Furthermore, these results can contribute to a broader understanding of magnetic field evolution in compact objects and offer new avenues for constraining the physical processes underlying magnetic field generation. Consequently, such approaches hold significant promise for astrophysical discovery, especially in parameter regimes where traditional observational techniques face limitations. Looking forward, this framework can be extended to larger WD samples from upcoming surveys (e.g., LSST and SDSS-VI) and more sophisticated magnetic field modeling incorporating temperature effects appropriately. These developments promise to significantly advance our understanding towards the origin of magnetic fields in degenerate stars and their roles in stellar astrophysics.

\section*{Acknowledgments} The authors gratefully acknowledge the anonymous reviewer for their comprehensive evaluation of the manuscript to improve its clarity, especially the methodological aspects. SK gratefully acknowledges the support and hospitality from ICTP, Italy during a research visit through the Associates Programme. SK and TB acknowledge funding from the National Science Centre, Poland (grant no. 2023/49/B/ST9/02777). AU and YM are supported by the National Key R\&D Program of China (grant no.\,2023YFE0101200), the National Natural Science Foundation of China (grant no.\,12273022 and 12511540053), the Research Fund for Excellent International PhD Students (grant no.\,W2442004) and the Shanghai Municipality orientation program of Basic Research for International Scientists (grant no.\,22JC1410600).

\section*{Data availability}
The WD data used in this article were accessed from \url{https://www.montrealwhitedwarfdatabase.org/home.html}. The data derived using the ML algorithm generated in this research will be shared upon reasonable request with the corresponding author.

\bibliographystyle{elsarticle-harv} 
\bibliography{Bibliography}

@ARTICLE{2018MNRAS.480.1547L,
       author = {{Lauffer}, G.~R. and {Romero}, A.~D. and {Kepler}, S.~O.},
        title = "{New full evolutionary sequences of H- and He-atmosphere massive white dwarf stars using MESA}",
      journal = {\mnras},
         year = 2018,
        month = oct,
       volume = {480},
       number = {2},
        pages = {1547-1562},
          doi = {10.1093/mnras/sty1925},
       adsurl = {https://ui.adsabs.harvard.edu/abs/2018MNRAS.480.1547L}
}

@ARTICLE{1935MNRAS..95..207C,
       author = {{Chandrasekhar}, S.},
        title = "{The highly collapsed configurations of a stellar mass (Second paper)}",
      journal = {\mnras},
         year = 1935,
        month = jan,
       volume = {95},
        pages = {207-225},
          doi = {10.1093/mnras/95.3.207},
       adsurl = {https://ui.adsabs.harvard.edu/abs/1935MNRAS..95..207C}
}

@ARTICLE{2006Natur.443..308H,
   author = {{Howell}, D.~A. and {Sullivan}, M. and {Nugent}, P.~E. and {Ellis}, R.~S. and 
	{Conley}, A.~J. and {Le Borgne}, D. and {Carlberg}, R.~G. and 
	{Guy}, J. and {Balam}, D. and {Basa}, S. and {Fouchez}, D. and 
	{Hook}, I.~M. and {Hsiao}, E.~Y. and {Neill}, J.~D. and {Pain}, R. and 
	{Perrett}, K.~M. and {Pritchet}, C.~J.},
    title = "{The type Ia supernova SNLS-03D3bb from a super-Chandrasekhar-mass white dwarf star}",
  journal = {\nat},
   eprint = {astro-ph/0609616},
     year = 2006,
    month = sep,
   volume = 443,
    pages = {308-311},
      doi = {10.1038/nature05103},
   adsurl = {http://adsabs.harvard.edu/abs/2006Natur.443..308H}
}

@ARTICLE{2010ApJ...713.1073S,
       author = {{Scalzo}, R.~A. and {Aldering}, G. and {Antilogus}, P. and {Aragon}, C. and {Bailey}, S. and {Baltay}, C. and {Bongard}, S. and {Buton}, C. and {Childress}, M. and {Chotard}, N. and {Copin}, Y. and {Fakhouri}, H.~K. and {Gal-Yam}, A. and {Gangler}, E. and {Hoyer}, S. and {Kasliwal}, M. and {Loken}, S. and {Nugent}, P. and {Pain}, R. and {P{\'e}contal}, E. and {Pereira}, R. and {Perlmutter}, S. and {Rabinowitz}, D. and {Rau}, A. and {Rigaudier}, G. and {Runge}, K. and {Smadja}, G. and {Tao}, C. and {Thomas}, R.~C. and {Weaver}, B. and {Wu}, C.},
        title = "{Nearby Supernova Factory Observations of SN 2007if: First Total Mass Measurement of a Super-Chandrasekhar-Mass Progenitor}",
      journal = {\apj},
         year = 2010,
        month = apr,
       volume = {713},
       number = {2},
        pages = {1073-1094},
          doi = {10.1088/0004-637X/713/2/1073},
archivePrefix = {arXiv},
       eprint = {1003.2217},
 primaryClass = {astro-ph.CO},
       adsurl = {https://ui.adsabs.harvard.edu/abs/2010ApJ...713.1073S}
}

@ARTICLE{2020ApJ...896...69K,
       author = {{Kalita}, Surajit and {Mukhopadhyay}, Banibrata and {Mondal}, Tushar and {Bulik}, Tomasz},
        title = "{Timescales for Detection of Super-Chandrasekhar White Dwarfs by Gravitational-wave Astronomy}",
      journal = {\apj},
         year = 2020,
        month = jun,
       volume = {896},
       number = {1},
          eid = {69},
        pages = {69},
          doi = {10.3847/1538-4357/ab8e40},
archivePrefix = {arXiv},
       eprint = {2004.13750},
 primaryClass = {astro-ph.HE},
       adsurl = {https://ui.adsabs.harvard.edu/abs/2020ApJ...896...69K}
}

@ARTICLE{2021MNRAS.508..842K,
       author = {{Kalita}, Surajit and {Mondal}, Tushar and {Tout}, Christopher A. and {Bulik}, Tomasz and {Mukhopadhyay}, Banibrata},
        title = "{Resolving dichotomy in compact objects through continuous gravitational waves observation}",
      journal = {\mnras},
         year = 2021,
        month = nov,
       volume = {508},
       number = {1},
        pages = {842-851},
          doi = {10.1093/mnras/stab2625},
archivePrefix = {arXiv},
       eprint = {2109.06246},
 primaryClass = {astro-ph.HE},
       adsurl = {https://ui.adsabs.harvard.edu/abs/2021MNRAS.508..842K}
}

@ARTICLE{2022PhRvD.105b4028S,
       author = {{Sarmah}, Lupamudra and {Kalita}, Surajit and {Wojnar}, Aneta},
        title = "{Stability criterion for white dwarfs in Palatini $f(R)$ gravity}",
      journal = {\prd},
         year = 2022,
        month = jan,
       volume = {105},
       number = {2},
          eid = {024028},
        pages = {024028},
          doi = {10.1103/PhysRevD.105.024028},
archivePrefix = {arXiv},
       eprint = {2111.08029},
 primaryClass = {gr-qc},
       adsurl = {https://ui.adsabs.harvard.edu/abs/2022PhRvD.105b4028S}
}

@ARTICLE{2024MNRAS.529.4577K,
       author = {{Karinkuzhi}, Drisya and {Mukhopadhyay}, Banibrata and {Wickramasinghe}, Dayal and {Tout}, Christopher A.},
        title = "{Mass-radius relation for magnetized white dwarfs from SDSS}",
      journal = {\mnras},
         year = 2024,
        month = apr,
       volume = {529},
       number = {4},
        pages = {4577-4584},
          doi = {10.1093/mnras/stae829},
archivePrefix = {arXiv},
       eprint = {2403.13888},
 primaryClass = {astro-ph.SR},
       adsurl = {https://ui.adsabs.harvard.edu/abs/2024MNRAS.529.4577K}
}

@ARTICLE{2013PhRvL.110g1102D,
       author = {{Das}, Upasana and {Mukhopadhyay}, Banibrata},
        title = "{New Mass Limit for White Dwarfs: Super-Chandrasekhar Type Ia Supernova as a New Standard Candle}",
      journal = {\prl},
         year = 2013,
        month = feb,
       volume = {110},
       number = {7},
          eid = {071102},
        pages = {071102},
          doi = {10.1103/PhysRevLett.110.071102},
archivePrefix = {arXiv},
       eprint = {1301.5965},
 primaryClass = {astro-ph.SR},
       adsurl = {https://ui.adsabs.harvard.edu/abs/2013PhRvL.110g1102D}
}

@ARTICLE{2024MNRAS.534L..65G,
       author = {{Ginzburg}, Sivan},
        title = "{Younger age for the oldest magnetic white dwarfs}",
      journal = {\mnras},
         year = 2024,
        month = oct,
       volume = {534},
       number = {1},
        pages = {L65-L70},
          doi = {10.1093/mnrasl/slae082},
archivePrefix = {arXiv},
       eprint = {2408.04695},
 primaryClass = {astro-ph.SR},
       adsurl = {https://ui.adsabs.harvard.edu/abs/2024MNRAS.534L..65G}
}

@ARTICLE{2015SSRv..191..111F,
       author = {{Ferrario}, Lilia and {de Martino}, Domitilla and {G{\"a}nsicke}, Boris T.},
        title = "{Magnetic White Dwarfs}",
      journal = {\ssr},
         year = 2015,
        month = oct,
       volume = {191},
       number = {1-4},
        pages = {111-169},
          doi = {10.1007/s11214-015-0152-0},
archivePrefix = {arXiv},
       eprint = {1504.08072},
 primaryClass = {astro-ph.SR},
       adsurl = {https://ui.adsabs.harvard.edu/abs/2015SSRv..191..111F}
}

@ARTICLE{2012A&A...545A..30L,
       author = {{Landstreet}, J.~D. and {Bagnulo}, S. and {Valyavin}, G.~G. and {Fossati}, L. and {Jordan}, S. and {Monin}, D. and {Wade}, G.~A.},
        title = "{On the incidence of weak magnetic fields in DA white dwarfs}",
      journal = {\aap},
         year = 2012,
        month = sep,
       volume = {545},
          eid = {A30},
        pages = {A30},
          doi = {10.1051/0004-6361/201219829},
archivePrefix = {arXiv},
       eprint = {1208.3650},
 primaryClass = {astro-ph.SR},
       adsurl = {https://ui.adsabs.harvard.edu/abs/2012A&A...545A..30L}
}

@ARTICLE{2023ApJ...944...56A,
       author = {{Amorim}, L.~L. and {Kepler}, S.~O. and {K{\"u}lebi}, Baybars and {Jordan}, S. and {Romero}, A.~D.},
        title = "{Catalog of Magnetic White Dwarfs with Hydrogen Dominated Atmospheres}",
      journal = {\apj},
         year = 2023,
        month = feb,
       volume = {944},
       number = {1},
          eid = {56},
        pages = {56},
          doi = {10.3847/1538-4357/acaf6e},
archivePrefix = {arXiv},
       eprint = {2301.08862},
 primaryClass = {astro-ph.SR},
       adsurl = {https://ui.adsabs.harvard.edu/abs/2023ApJ...944...56A}
}

@ARTICLE{2007ApJ...654..499K,
       author = {{Kawka}, A. and {Vennes}, S. and {Schmidt}, G.~D. and {Wickramasinghe}, D.~T. and {Koch}, R.},
        title = "{Spectropolarimetric Survey of Hydrogen-rich White Dwarf Stars}",
      journal = {\apj},
         year = 2007,
        month = jan,
       volume = {654},
       number = {1},
        pages = {499-520},
          doi = {10.1086/509072},
archivePrefix = {arXiv},
       eprint = {astro-ph/0609273},
 primaryClass = {astro-ph},
       adsurl = {https://ui.adsabs.harvard.edu/abs/2007ApJ...654..499K}
}

@ARTICLE{2000MNRAS.317..310H,
       author = {{Heyl}, Jeremy S.},
        title = "{Gravitational radiation from strongly magnetized white dwarfs}",
      journal = {\mnras},
         year = 2000,
        month = sep,
       volume = {317},
       number = {2},
        pages = {310-314},
          doi = {10.1046/j.1365-8711.2000.03533.x},
archivePrefix = {arXiv},
       eprint = {astro-ph/0001343},
 primaryClass = {astro-ph},
       adsurl = {https://ui.adsabs.harvard.edu/abs/2000MNRAS.317..310H}
}

@ARTICLE{2014MNRAS.437..675W,
       author = {{Wickramasinghe}, Dayal T. and {Tout}, Christopher A. and {Ferrario}, Lilia},
        title = "{The most magnetic stars}",
      journal = {\mnras},
         year = 2014,
        month = jan,
       volume = {437},
       number = {1},
        pages = {675-681},
          doi = {10.1093/mnras/stt1910},
archivePrefix = {arXiv},
       eprint = {1310.2696},
 primaryClass = {astro-ph.SR},
       adsurl = {https://ui.adsabs.harvard.edu/abs/2014MNRAS.437..675W}
}

@ARTICLE{2019MNRAS.490.2692K,
       author = {{Kalita}, Surajit and {Mukhopadhyay}, Banibrata},
        title = "{Continuous gravitational wave from magnetized white dwarfs and neutron stars: possible missions for LISA, DECIGO, BBO, ET detectors}",
      journal = {\mnras},
         year = 2019,
        month = dec,
       volume = {490},
       number = {2},
        pages = {2692-2705},
          doi = {10.1093/mnras/stz2734},
archivePrefix = {arXiv},
       eprint = {1905.02730},
 primaryClass = {astro-ph.HE},
       adsurl = {https://ui.adsabs.harvard.edu/abs/2019MNRAS.490.2692K}
}

@INPROCEEDINGS{2017ASPC..509....3D,
       author = {{Dufour}, P. and {Blouin}, S. and {Coutu}, S. and {Fortin-Archambault}, M. and {Thibeault}, C. and {Bergeron}, P. and {Fontaine}, G.},
        title = "{The Montreal White Dwarf Database: A Tool for the Community}",
    booktitle = {20th European White Dwarf Workshop},
         year = 2017,
       editor = {{Tremblay}, P. -E. and {Gaensicke}, B. and {Marsh}, T.},
       series = {Astronomical Society of the Pacific Conference Series},
       volume = {509},
        month = mar,
        pages = {3},
          doi = {10.48550/arXiv.1610.00986},
archivePrefix = {arXiv},
       eprint = {1610.00986},
 primaryClass = {astro-ph.SR},
       adsurl = {https://ui.adsabs.harvard.edu/abs/2017ASPC..509....3D}
}

@ARTICLE{1981ApJS...45..457A,
       author = {{Angel}, J.~R.~P. and {Borra}, E.~F. and {Landstreet}, J.~D.},
        title = "{The magnetic fields of white dwarfs.}",
      journal = {\apjs},
         year = 1981,
        month = mar,
       volume = {45},
        pages = {457-474},
          doi = {10.1086/190720},
       adsurl = {https://ui.adsabs.harvard.edu/abs/1981ApJS...45..457A}
}

@ARTICLE{2008MNRAS.387..897T,
       author = {{Tout}, C.~A. and {Wickramasinghe}, D.~T. and {Liebert}, J. and {Ferrario}, L. and {Pringle}, J.~E.},
        title = "{Binary star origin of high field magnetic white dwarfs}",
      journal = {\mnras},
         year = 2008,
        month = jun,
       volume = {387},
       number = {2},
        pages = {897-901},
          doi = {10.1111/j.1365-2966.2008.13291.x},
archivePrefix = {arXiv},
       eprint = {0805.0115},
 primaryClass = {astro-ph},
       adsurl = {https://ui.adsabs.harvard.edu/abs/2008MNRAS.387..897T}
}

@ARTICLE{2012ApJ...749...25G,
       author = {{Garc{\'\i}a-Berro}, Enrique and {Lor{\'e}n-Aguilar}, Pablo and {Aznar-Sigu{\'a}n}, Gabriela and {Torres}, Santiago and {Camacho}, Judit and {Althaus}, Leandro G. and {C{\'o}rsico}, Alejandro H. and {K{\"u}lebi}, Baybars and {Isern}, Jordi},
        title = "{Double Degenerate Mergers as Progenitors of High-field Magnetic White Dwarfs}",
      journal = {\apj},
         year = 2012,
        month = apr,
       volume = {749},
       number = {1},
          eid = {25},
        pages = {25},
          doi = {10.1088/0004-637X/749/1/25},
archivePrefix = {arXiv},
       eprint = {1202.0461},
 primaryClass = {astro-ph.SR},
       adsurl = {https://ui.adsabs.harvard.edu/abs/2012ApJ...749...25G}
}

@ARTICLE{2015MNRAS.447.1713B,
       author = {{Briggs}, Gordon P. and {Ferrario}, Lilia and {Tout}, Christopher A. and {Wickramasinghe}, Dayal T. and {Hurley}, Jarrod R.},
        title = "{Merging binary stars and the magnetic white dwarfs}",
      journal = {\mnras},
         year = 2015,
        month = feb,
       volume = {447},
       number = {2},
        pages = {1713-1723},
          doi = {10.1093/mnras/stu2539},
archivePrefix = {arXiv},
       eprint = {1412.5662},
 primaryClass = {astro-ph.SR},
       adsurl = {https://ui.adsabs.harvard.edu/abs/2015MNRAS.447.1713B}
}

@ARTICLE{2013MNRAS.429.2934K,
       author = {{Kepler}, S.~O. and {Pelisoli}, I. and {Jordan}, S. and {Kleinman}, S.~J. and {Koester}, D. and {K{\"u}lebi}, B. and {Pe{\c{c}}anha}, V. and {Castanheira}, B.~G. and {Nitta}, A. and {Costa}, J.~E.~S. and {Winget}, D.~E. and {Kanaan}, A. and {Fraga}, L.},
        title = "{Magnetic white dwarf stars in the Sloan Digital Sky Survey}",
      journal = {\mnras},
         year = 2013,
        month = mar,
       volume = {429},
       number = {4},
        pages = {2934-2944},
          doi = {10.1093/mnras/sts522},
archivePrefix = {arXiv},
       eprint = {1211.5709},
 primaryClass = {astro-ph.SR},
       adsurl = {https://ui.adsabs.harvard.edu/abs/2013MNRAS.429.2934K}
}

@BOOK{2014sdmm.book.....I,
       author = {{Ivezi{\'c}}, {\v{Z}}eljko and {Connolly}, Andrew J. and {VanderPlas}, Jacob T. and {Gray}, Alexander},
        title = "{Statistics, Data Mining, and Machine Learning in Astronomy: A Practical Python Guide for the Analysis of Survey Data}",
         year = 2014,
    publisher = {Princeton University Press},
         ISBN = {9780691151687},
          doi = {10.1515/9781400848911},
       adsurl = {https://ui.adsabs.harvard.edu/abs/2014sdmm.book.....I}
}

@ARTICLE{2021A&A...648A..74H,
       author = {{Heydenreich}, Sven and {Br{\"u}ck}, Benjamin and {Harnois-D{\'e}raps}, Joachim},
        title = "{Persistent homology in cosmic shear: Constraining parameters with topological data analysis}",
      journal = {\aap},
         year = 2021,
        month = apr,
       volume = {648},
          eid = {A74},
        pages = {A74},
          doi = {10.1051/0004-6361/202039048},
archivePrefix = {arXiv},
       eprint = {2007.13724},
 primaryClass = {astro-ph.CO},
       adsurl = {https://ui.adsabs.harvard.edu/abs/2021A&A...648A..74H}
}

@ARTICLE{2023A&A...671A..19R,
       author = {{Rosito}, M.~S. and {Bignone}, L.~A. and {Tissera}, P.~B. and {Pedrosa}, S.~E.},
        title = "{Application of dimensionality reduction and clustering algorithms for the classification of kinematic morphologies of galaxies}",
      journal = {\aap},
         year = 2023,
        month = mar,
       volume = {671},
          eid = {A19},
        pages = {A19},
          doi = {10.1051/0004-6361/202244707},
archivePrefix = {arXiv},
       eprint = {2212.03999},
 primaryClass = {astro-ph.GA},
       adsurl = {https://ui.adsabs.harvard.edu/abs/2023A&A...671A..19R}
}

@ARTICLE{2022PhRvD.106b3017C,
       author = {{Christian}, Pierre and {Chan}, Chi-kwan and {Hsu}, Anthony and {{\"O}zel}, Feryal and {Psaltis}, Dimitrios and {Natarajan}, Iniyan},
        title = "{Topological data analysis of black hole images}",
      journal = {\prd},
         year = 2022,
        month = jul,
       volume = {106},
       number = {2},
          eid = {023017},
        pages = {023017},
          doi = {10.1103/PhysRevD.106.023017},
archivePrefix = {arXiv},
       eprint = {2202.00698},
 primaryClass = {astro-ph.HE},
       adsurl = {https://ui.adsabs.harvard.edu/abs/2022PhRvD.106b3017C}
}

@ARTICLE{2016ApJ...820....8S,
       author = {{Saz Parkinson}, P.~M. and {Xu}, H. and {Yu}, P.~L.~H. and {Salvetti}, D. and {Marelli}, M. and {Falcone}, A.~D.},
        title = "{Classification and Ranking of Fermi LAT Gamma-ray Sources from the 3FGL Catalog using Machine Learning Techniques}",
      journal = {\apj},
         year = 2016,
        month = mar,
       volume = {820},
       number = {1},
          eid = {8},
        pages = {8},
          doi = {10.3847/0004-637X/820/1/8},
archivePrefix = {arXiv},
       eprint = {1602.00385},
 primaryClass = {astro-ph.HE},
       adsurl = {https://ui.adsabs.harvard.edu/abs/2016ApJ...820....8S}
}

@ARTICLE{2018arXiv180203426M,
       author = {{McInnes}, Leland and {Healy}, John and {Melville}, James},
        title = "{UMAP: Uniform Manifold Approximation and Projection for Dimension Reduction}",
      journal = {arXiv e-prints},
         year = 2018,
        month = feb,
          eid = {arXiv:1802.03426},
        pages = {arXiv:1802.03426},
          doi = {10.48550/arXiv.1802.03426},
archivePrefix = {arXiv},
       eprint = {1802.03426},
 primaryClass = {stat.ML},
       adsurl = {https://ui.adsabs.harvard.edu/abs/2018arXiv180203426M}
}

@INPROCEEDINGS{1996kddm.conf..226E,
       author = {{Ester}, Martin and {Kriegel}, Hans-Peter and {Sander}, J{\"o}rg and {Xu}, Xiaowei},
        title = "{A Density-Based Algorithm for Discovering Clusters in Large Spatial Databases with Noise}",
    booktitle = {Second International Conference on Knowledge Discovery and Data Mining (KDD'96). Proceedings of a conference held August 2-4},
         year = 1996,
       editor = {{Pfitzner}, D.~W. and {Salmon}, J.~K.},
        month = jan,
        pages = {226-331},
       adsurl = {https://ui.adsabs.harvard.edu/abs/1996kddm.conf..226E}
}

@ARTICLE{2006Sci...313..504H,
       author = {{Hinton}, G.~E. and {Salakhutdinov}, R.~R.},
        title = "{Reducing the Dimensionality of Data with Neural Networks}",
      journal = {Science},
         year = 2006,
        month = jul,
       volume = {313},
       number = {5786},
        pages = {504-507},
          doi = {10.1126/science.1127647},
       adsurl = {https://ui.adsabs.harvard.edu/abs/2006Sci...313..504H}
}

@ARTICLE{1992ApJ...394..228B,
       author = {{Bergeron}, P. and {Saffer}, Rex A. and {Liebert}, James},
        title = "{A Spectroscopic Determination of the Mass Distribution of DA White Dwarfs}",
      journal = {\apj},
         year = 1992,
        month = jul,
       volume = {394},
        pages = {228},
          doi = {10.1086/171575},
       adsurl = {https://ui.adsabs.harvard.edu/abs/1992ApJ...394..228B}
}

@ARTICLE{2009ApJ...696.1755T,
       author = {{Tremblay}, P. -E. and {Bergeron}, P.},
        title = "{Spectroscopic Analysis of DA White Dwarfs: Stark Broadening of Hydrogen Lines Including Nonideal Effects}",
      journal = {\apj},
         year = 2009,
        month = may,
       volume = {696},
       number = {2},
        pages = {1755-1770},
          doi = {10.1088/0004-637X/696/2/1755},
archivePrefix = {arXiv},
       eprint = {0902.4182},
 primaryClass = {astro-ph.SR},
       adsurl = {https://ui.adsabs.harvard.edu/abs/2009ApJ...696.1755T}
}

@ARTICLE{2011ApJ...743..138G,
       author = {{Gianninas}, A. and {Bergeron}, P. and {Ruiz}, M.~T.},
        title = "{A Spectroscopic Survey and Analysis of Bright, Hydrogen-rich White Dwarfs}",
      journal = {\apj},
         year = 2011,
        month = dec,
       volume = {743},
       number = {2},
          eid = {138},
        pages = {138},
          doi = {10.1088/0004-637X/743/2/138},
archivePrefix = {arXiv},
       eprint = {1109.3171},
 primaryClass = {astro-ph.SR},
       adsurl = {https://ui.adsabs.harvard.edu/abs/2011ApJ...743..138G}
}

@ARTICLE{2006AJ....132.1221H,
       author = {{Holberg}, J.~B. and {Bergeron}, Pierre},
        title = "{Calibration of Synthetic Photometry Using DA White Dwarfs}",
      journal = {\aj},
         year = 2006,
        month = sep,
       volume = {132},
       number = {3},
        pages = {1221-1233},
          doi = {10.1086/505938},
       adsurl = {https://ui.adsabs.harvard.edu/abs/2006AJ....132.1221H}
}

@ARTICLE{2019ApJ...871..169G,
       author = {{Genest-Beaulieu}, C. and {Bergeron}, P.},
        title = "{A Comprehensive Spectroscopic and Photometric Analysis of DA and DB White Dwarfs from SDSS and Gaia}",
      journal = {\apj},
         year = 2019,
        month = feb,
       volume = {871},
       number = {2},
          eid = {169},
        pages = {169},
          doi = {10.3847/1538-4357/aafac6},
       adsurl = {https://ui.adsabs.harvard.edu/abs/2019ApJ...871..169G}
}

@ARTICLE{2001PASP..113..409F,
       author = {{Fontaine}, G. and {Brassard}, P. and {Bergeron}, P.},
        title = "{The Potential of White Dwarf Cosmochronology}",
      journal = {\pasp},
         year = 2001,
        month = apr,
       volume = {113},
       number = {782},
        pages = {409-435},
          doi = {10.1086/319535},
       adsurl = {https://ui.adsabs.harvard.edu/abs/2001PASP..113..409F}
}

@ARTICLE{2010A&ARv..18..471A,
       author = {{Althaus}, Leandro G. and {C{\'o}rsico}, Alejandro H. and {Isern}, Jordi and {Garc{\'\i}a-Berro}, Enrique},
        title = "{Evolutionary and pulsational properties of white dwarf stars}",
      journal = {\aapr},
         year = 2010,
        month = oct,
       volume = {18},
       number = {4},
        pages = {471-566},
          doi = {10.1007/s00159-010-0033-1},
archivePrefix = {arXiv},
       eprint = {1007.2659},
 primaryClass = {astro-ph.SR},
       adsurl = {https://ui.adsabs.harvard.edu/abs/2010A&ARv..18..471A}
}

@ARTICLE{2017MNRAS.465.2849T,
       author = {{Tremblay}, P. -E. and {Gentile-Fusillo}, N. and {Raddi}, R. and {Jordan}, S. and {Besson}, C. and {G{\"a}nsicke}, B.~T. and {Parsons}, S.~G. and {Koester}, D. and {Marsh}, T. and {Bohlin}, R. and {Kalirai}, J. and {Deustua}, S.},
        title = "{The Gaia DR1 mass-radius relation for white dwarfs}",
      journal = {\mnras},
         year = 2017,
        month = mar,
       volume = {465},
       number = {3},
        pages = {2849-2861},
          doi = {10.1093/mnras/stw2854},
archivePrefix = {arXiv},
       eprint = {1611.00629},
 primaryClass = {astro-ph.SR},
       adsurl = {https://ui.adsabs.harvard.edu/abs/2017MNRAS.465.2849T}
}

@ARTICLE{2019MNRAS.482.4570G,
       author = {{Gentile Fusillo}, Nicola Pietro and {Tremblay}, Pier-Emmanuel and {G{\"a}nsicke}, Boris T. and {Manser}, Christopher J. and {Cunningham}, Tim and {Cukanovaite}, Elena and {Hollands}, Mark and {Marsh}, Thomas and {Raddi}, Roberto and {Jordan}, Stefan and {Toonen}, Silvia and {Geier}, Stephan and {Barstow}, Martin and {Cummings}, Jeffrey D.},
        title = "{A Gaia Data Release 2 catalogue of white dwarfs and a comparison with SDSS}",
      journal = {\mnras},
         year = 2019,
        month = feb,
       volume = {482},
       number = {4},
        pages = {4570-4591},
          doi = {10.1093/mnras/sty3016},
archivePrefix = {arXiv},
       eprint = {1807.03315},
 primaryClass = {astro-ph.SR},
       adsurl = {https://ui.adsabs.harvard.edu/abs/2019MNRAS.482.4570G}
}

@INCOLLECTION{1995LNP...443...41W,
       author = {{Wood}, Matt A.},
        title = "{Theoretical White Dwarf Luminosity Functions: DA Models}",
    booktitle = {White Dwarfs},
    publisher = "{Springer Berlin Heidelberg}",
      address = "{Berlin, Heidelberg}",
         year = 1995,
       editor = {{Koester}, Detlev and {Werner}, Klaus},
       volume = {443},
        pages = {41},
         isbn = "978-3-540-49202-3",
          doi = {10.1007/3-540-59157-5_171},
       adsurl = {https://ui.adsabs.harvard.edu/abs/1995LNP...443...41W}
}

@ARTICLE{2008ApJ...676..594K,
       author = {{Kalirai}, Jasonjot S. and {Hansen}, Brad M.~S. and {Kelson}, Daniel D. and {Reitzel}, David B. and {Rich}, R. Michael and {Richer}, Harvey B.},
        title = "{The Initial-Final Mass Relation: Direct Constraints at the Low-Mass End}",
      journal = {\apj},
         year = 2008,
        month = mar,
       volume = {676},
       number = {1},
        pages = {594-609},
          doi = {10.1086/527028},
archivePrefix = {arXiv},
       eprint = {0706.3894},
 primaryClass = {astro-ph},
       adsurl = {https://ui.adsabs.harvard.edu/abs/2008ApJ...676..594K}
}

@ARTICLE{2018ApJ...866...21C,
       author = {{Cummings}, Jeffrey D. and {Kalirai}, Jason S. and {Tremblay}, P. -E. and {Ramirez-Ruiz}, Enrico and {Choi}, Jieun},
        title = "{The White Dwarf Initial-Final Mass Relation for Progenitor Stars from 0.85 to 7.5 M $_{{\ensuremath{\odot}}}$}",
      journal = {\apj},
         year = 2018,
        month = oct,
       volume = {866},
       number = {1},
          eid = {21},
        pages = {21},
          doi = {10.3847/1538-4357/aadfd6},
archivePrefix = {arXiv},
       eprint = {1809.01673},
 primaryClass = {astro-ph.SR},
       adsurl = {https://ui.adsabs.harvard.edu/abs/2018ApJ...866...21C}
}

@ARTICLE{2023MNRAS.520.6111H,
       author = {{Hardy}, Fran{\c{c}}ois and {Dufour}, Patrick and {Jordan}, Stefan},
        title = "{Spectrophotometric analysis of magnetic white dwarf - I. Hydrogen-rich compositions}",
      journal = {\mnras},
         year = 2023,
        month = apr,
       volume = {520},
       number = {4},
        pages = {6111-6134},
          doi = {10.1093/mnras/stad196},
archivePrefix = {arXiv},
       eprint = {2301.06596},
 primaryClass = {astro-ph.SR},
       adsurl = {https://ui.adsabs.harvard.edu/abs/2023MNRAS.520.6111H}
}

@ARTICLE{2015ApJ...812...19T,
       author = {{Tremblay}, P. -E. and {Fontaine}, G. and {Freytag}, B. and {Steiner}, O. and {Ludwig}, H. -G. and {Steffen}, M. and {Wedemeyer}, S. and {Brassard}, P.},
        title = "{On the Evolution of Magnetic White Dwarfs}",
      journal = {\apj},
         year = 2015,
        month = oct,
       volume = {812},
       number = {1},
          eid = {19},
        pages = {19},
          doi = {10.1088/0004-637X/812/1/19},
archivePrefix = {arXiv},
       eprint = {1509.05398},
 primaryClass = {astro-ph.SR},
       adsurl = {https://ui.adsabs.harvard.edu/abs/2015ApJ...812...19T}
}

@ARTICLE{2002MNRAS.333..589C,
       author = {{Cumming}, Andrew},
        title = "{Magnetic field evolution in accreting white dwarfs}",
      journal = {\mnras},
         year = 2002,
        month = jul,
       volume = {333},
       number = {3},
        pages = {589-602},
          doi = {10.1046/j.1365-8711.2002.05434.x},
archivePrefix = {arXiv},
       eprint = {astro-ph/0202079},
 primaryClass = {astro-ph},
       adsurl = {https://ui.adsabs.harvard.edu/abs/2002MNRAS.333..589C}
}

@INPROCEEDINGS{2017ASPC..509..409I,
       author = {{Isern}, J. and {Garc{\'\i}a-Berro}, E. and {K{\"u}lebi}, B. and {Lor{\'e}n-Aguilar}, P.},
        title = "{Magnetic Fields and the Crystallization of White Dwarfs}",
    booktitle = {20th European White Dwarf Workshop},
         year = 2017,
       editor = {{Tremblay}, P. -E. and {Gaensicke}, B. and {Marsh}, T.},
       series = {Astronomical Society of the Pacific Conference Series},
       volume = {509},
        month = mar,
        pages = {409},
       adsurl = {https://ui.adsabs.harvard.edu/abs/2017ASPC..509..409I}
}

@ARTICLE{2024A&A...691L..21C,
       author = {{Camisassa}, M. and {Fuentes}, J.~R. and {Schreiber}, M.~R. and {Rebassa-Mansergas}, A. and {Torres}, S. and {Raddi}, R. and {Dominguez}, I.},
        title = "{Main sequence dynamo magnetic fields emerging in the white dwarf phase}",
      journal = {\aap},
         year = 2024,
        month = nov,
       volume = {691},
          eid = {L21},
        pages = {L21},
          doi = {10.1051/0004-6361/202452539},
archivePrefix = {arXiv},
       eprint = {2411.02296},
 primaryClass = {astro-ph.SR},
       adsurl = {https://ui.adsabs.harvard.edu/abs/2024A&A...691L..21C}
}

@ARTICLE{2024ApJ...970..181K,
       author = {{Kao}, Malia L. and {Hawkins}, Keith and {Rogers}, Laura K. and {Bonsor}, Amy and {Dunlap}, Bart H. and {Sanders}, Jason L. and {Montgomery}, M.~H. and {Winget}, D.~E.},
        title = "{Hunting for Polluted White Dwarfs and Other Treasures with Gaia XP Spectra and Unsupervised Machine Learning}",
      journal = {\apj},
         year = 2024,
        month = aug,
       volume = {970},
       number = {2},
          eid = {181},
        pages = {181},
          doi = {10.3847/1538-4357/ad5d6e},
archivePrefix = {arXiv},
       eprint = {2405.17667},
 primaryClass = {astro-ph.SR},
       adsurl = {https://ui.adsabs.harvard.edu/abs/2024ApJ...970..181K}
}

@ARTICLE{2024MNRAS.535.2246B,
       author = {{Byrne}, Xander and {Bonsor}, Amy and {Rogers}, Laura K. and {Manser}, Christopher J.},
        title = "{Semi-supervised spectral classification of DESI white dwarfs by dimensionality reduction}",
      journal = {\mnras},
         year = 2024,
        month = dec,
       volume = {535},
       number = {3},
        pages = {2246-2259},
          doi = {10.1093/mnras/stae2478},
archivePrefix = {arXiv},
       eprint = {2410.22221},
 primaryClass = {astro-ph.IM},
       adsurl = {https://ui.adsabs.harvard.edu/abs/2024MNRAS.535.2246B}
}

@ARTICLE{2024ApJ...977...31P,
       author = {{P{\'e}rez-Couto}, Xabier and {Pallas-Quintela}, Lara and {Manteiga}, Minia and {Villaver}, Eva and {Dafonte}, Carlos},
        title = "{Identifying New High-confidence Polluted White Dwarf Candidates Using Gaia XP Spectra and Self-organizing Maps}",
      journal = {\apj},
         year = 2024,
        month = dec,
       volume = {977},
       number = {1},
          eid = {31},
        pages = {31},
          doi = {10.3847/1538-4357/ad88f5},
archivePrefix = {arXiv},
       eprint = {2410.16015},
 primaryClass = {astro-ph.SR},
       adsurl = {https://ui.adsabs.harvard.edu/abs/2024ApJ...977...31P}
}

@ARTICLE{2025A&A...699A...3G,
       author = {{Garc{\'\i}a-Zamora}, Enrique Miguel and {Torres}, Santiago and {Rebassa-Mansergas}, Alberto and {Ferrer-Burjachs}, Aina},
        title = "{A random forest spectral classification of the Gaia 500 pc white dwarf population}",
      journal = {\aap},
         year = 2025,
        month = jun,
       volume = {699},
          eid = {A3},
        pages = {A3},
          doi = {10.1051/0004-6361/202554414},
archivePrefix = {arXiv},
       eprint = {2505.05560},
 primaryClass = {astro-ph.SR},
       adsurl = {https://ui.adsabs.harvard.edu/abs/2025A&A...699A...3G}
}

@ARTICLE{2025ApJ...988...51P,
       author = {{P{\'e}rez-Couto}, Xabier and {Manteiga}, Minia and {Villaver}, Eva},
        title = "{Finding White Dwarfs' Hidden Companions Using an Unsupervised Machine Learning Technique}",
      journal = {\apj},
         year = 2025,
        month = jul,
       volume = {988},
       number = {1},
          eid = {51},
        pages = {51},
          doi = {10.3847/1538-4357/addfd7},
archivePrefix = {arXiv},
       eprint = {2503.04672},
 primaryClass = {astro-ph.SR},
       adsurl = {https://ui.adsabs.harvard.edu/abs/2025ApJ...988...51P}
}

@ARTICLE{2025A&A...701A.273L,
       author = {{L{\'o}pez-Sanjuan}, C. and {Tremblay}, P.-E. and {del Pino}, A. and {Dom{\'\i}nguez S{\'a}nchez}, H. and {V{\'a}zquez Rami{\'o}}, H. and {Ederoclite}, A. and {Cenarro}, A.~J. and {Mar{\'\i}n-Franch}, A. and {Anguiano}, B. and {Civera}, T. and {Cruz}, P. and {Fern{\'a}ndez-Ontiveros}, J.~A. and {Jim{\'e}nez-Esteban}, F.~M. and {Rebassa-Mansergas}, A. and {Vega-Ferrero}, J. and {Alcaniz}, J. and {Angulo}, R.~E. and {Crist{\'o}bal-Hornillos}, D. and {Dupke}, R.~A. and {Hern{\'a}ndez-Monteagudo}, C. and {Moles}, M. and {Sodr{\'e}}, Jr., L. and {Varela}, J.},
        title = "{J-PLUS: Understanding outlier white dwarfs in the third data release via dimensionality reduction}",
      journal = {\aap},
         year = 2025,
        month = sep,
       volume = {701},
          eid = {A273},
        pages = {A273},
          doi = {10.1051/0004-6361/202555286},
archivePrefix = {arXiv},
       eprint = {2509.21304},
 primaryClass = {astro-ph.SR},
       adsurl = {https://ui.adsabs.harvard.edu/abs/2025A&A...701A.273L}
}

@ARTICLE{2025MNRAS.544.1939V,
       author = {{Vincent}, Olivier and {Dufour}, Patrick and {Bergeron}, Pierre},
        title = "{Neural posterior estimation for white dwarf spectroscopic characterization}",
      journal = {\mnras},
         year = 2025,
        month = dec,
       volume = {544},
       number = {2},
        pages = {1939-1949},
          doi = {10.1093/mnras/staf1820},
archivePrefix = {arXiv},
       eprint = {2510.16261},
 primaryClass = {astro-ph.SR},
       adsurl = {https://ui.adsabs.harvard.edu/abs/2025MNRAS.544.1939V}
}

@ARTICLE{2025ApJ...982...16Q,
       author = {{Qiang}, Da-Chun and {Zheng}, Jie and {You}, Zhi-Qiang and {Yang}, Sheng},
        title = "{Unsupervised Machine Learning for Classifying CHIME Fast Radio Bursts and Investigating Empirical Relations}",
      journal = {\apj},
         year = 2025,
        month = mar,
       volume = {982},
       number = {1},
          eid = {16},
        pages = {16},
          doi = {10.3847/1538-4357/adb72b},
archivePrefix = {arXiv},
       eprint = {2411.14040},
 primaryClass = {astro-ph.HE},
       adsurl = {https://ui.adsabs.harvard.edu/abs/2025ApJ...982...16Q}
}

@ARTICLE{2026JHEAp..4900449J,
       author = {{J{\'u}nior}, Ailton J.~B. and {Fortunato}, J{\'e}ferson A.~S. and {Silvestre}, Leonardo J. and {Alencar}, Thonimar V. and {Hip{\'o}lito-Ricaldi}, Wiliam S.},
        title = "{Comparative analysis of machine learning techniques for feature selection and classification of Fast Radio Bursts}",
      journal = {Journal of High Energy Astrophysics},
         year = 2026,
        month = jan,
       volume = {49},
          eid = {100449},
        pages = {100449},
          doi = {10.1016/j.jheap.2025.100449},
archivePrefix = {arXiv},
       eprint = {2506.18854},
 primaryClass = {astro-ph.HE},
       adsurl = {https://ui.adsabs.harvard.edu/abs/2026JHEAp..4900449J}
}


\appendix

\section{Machine learning methodology details}\label{Appendix}

\subsection{Dimensionality Reduction with UMAP}

\begin{figure}[htpb]
    \centering
    \subfigure[\texttt{n\_neighbors} = 5, \texttt{min\_dist} = 0.1]{\includegraphics[scale=0.5]{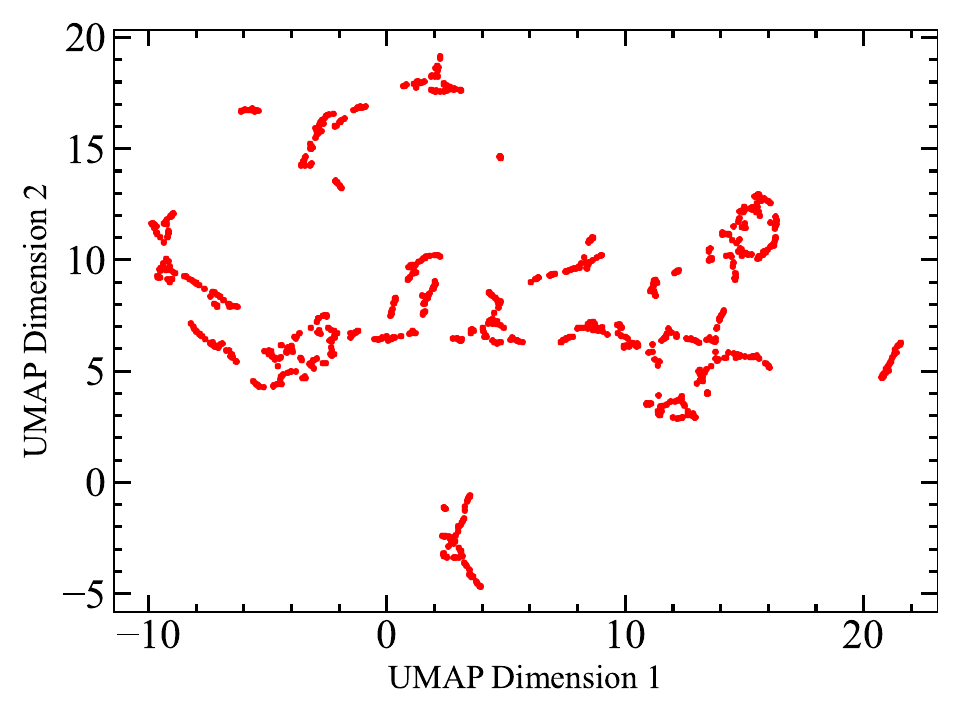}}
	\subfigure[\texttt{n\_neighbors} = 15, \texttt{min\_dist} = 0.3]{\includegraphics[scale=0.5]{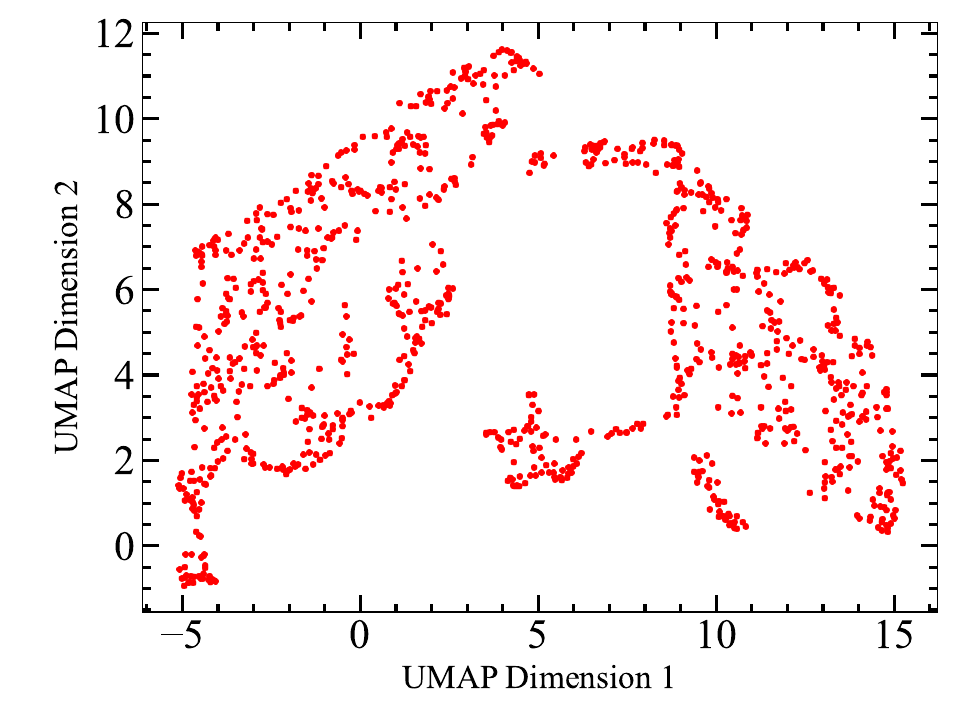}}
    \caption{UMAP dimensionality reduction with different hyperparameters.}
    \label{Fig: hyperparameters}
\end{figure}

We implement the UMAP technique to reduce the 6-dimensional WD parameter space (mass, surface gravity, luminosity, cooling age, effective temperature, and absolute magnitude) to a 2-dimensional space optimized for both visualization and neighborhood analysis. UMAP first constructs a high-dimensional weighted graph whose topology corresponds to the topological structure of the input data and then uses stochastic gradient descent to optimize a low-dimensional representation that preserves both local and global relationships among input data. The algorithm requires two key parameters. The first is \texttt{n\_neighbors}, which determines the balance between local and global structure preservation, where larger values correspond to broader trends whereas smaller values capture finer-scale variations. We use \texttt{n\_neighbors}=15 to optimally balance both local clustering with broader trends. The second parameter, \texttt{min\_dist}, sets the minimum distance between points in the low-dimensional space, affecting cluster compactness. We adopt \texttt{min\_dist}=0.1 to optimally retain some spread while keeping similar objects relatively close. Smaller values of these parameters would produce fragmented structures, while larger values would cause clusters to converge and eventually become indistinguishable from each other. A comparative visualization is provided in Fig.~\ref{Fig: hyperparameters} to illustrate these effects. Notably, moderate variations in the parameter range (\texttt{n\_neighbors} $\approx$ 10--20 and \texttt{min\_dist} $\approx$ 0.1--0.2) will not make any significant difference in the resulting embedding, demonstrating the overall stability of UMAP algorithm over other methods like t-SNE.

Before performing dimensionality reduction, all features were standardized using Python's \texttt{StandardScaler} module from \texttt{sklearn.preprocessing} so that it ensures equal weighting in the distance metric. It is applicable because the majority of WD parameters follow approximately Gaussian-like distributions after cleaning. One can, in principle, use \texttt{RobustScaler} instead, which scales using the median and interquartile range. While the detailed geometry of the 2D UMAP embedding would be altered, the astrophysically relevant structures remained unchanged. In particular, magnetized WDs continue to group together in one cluster and remain separated from the non-magnetic population, along with the high field ones stay in one branch, as obtained in the \texttt{StandardScaler} case. The resulting 2-dimensional space preserves the fundamental topological relations required for subsequent density-based clustering and regression tasks. 

\begin{figure}[htpb]
    \centering
    \subfigure[High correlated data (Correlation coefficient = 0.997).]{%
        \begin{minipage}{\linewidth}
            \centering
            \includegraphics[scale=0.5]{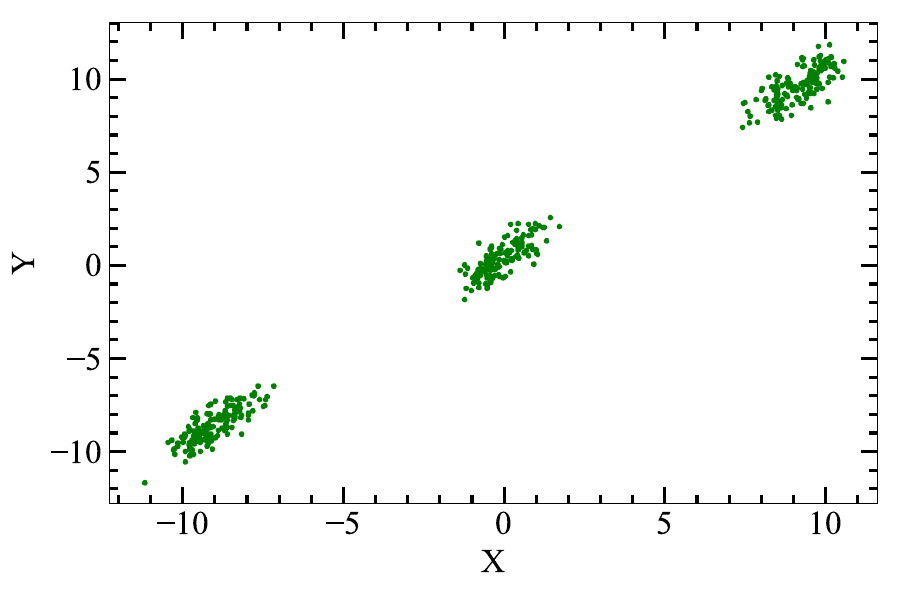}
            
            \includegraphics[scale=0.5]{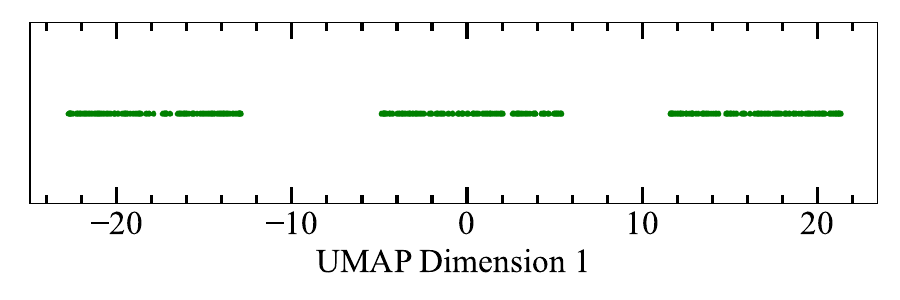}
        \end{minipage}
    }
    
    \subfigure[Low correlated data (Correlation coefficient = 0.322).]{%
        \begin{minipage}{\linewidth}
            \centering
            \includegraphics[scale=0.5]{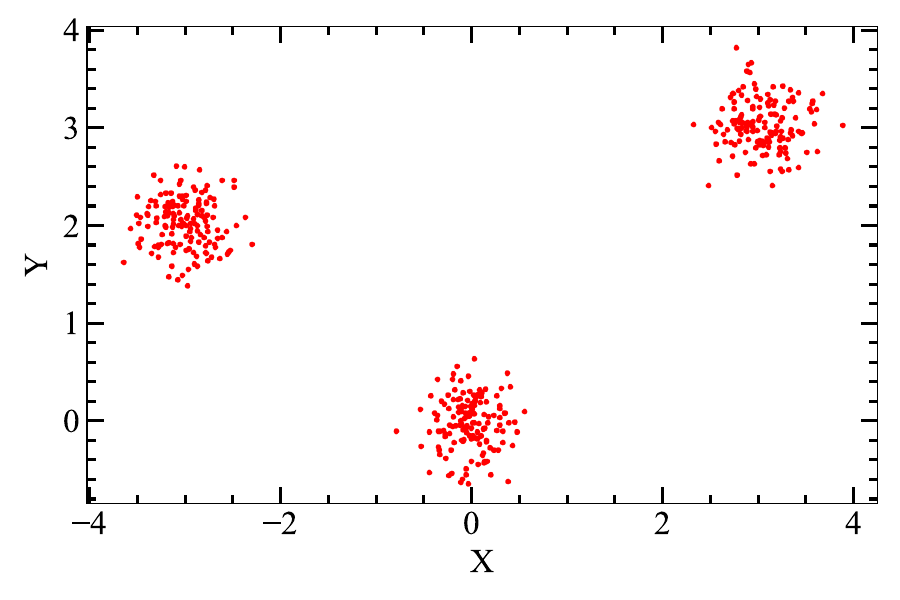}
            
            \includegraphics[scale=0.5]{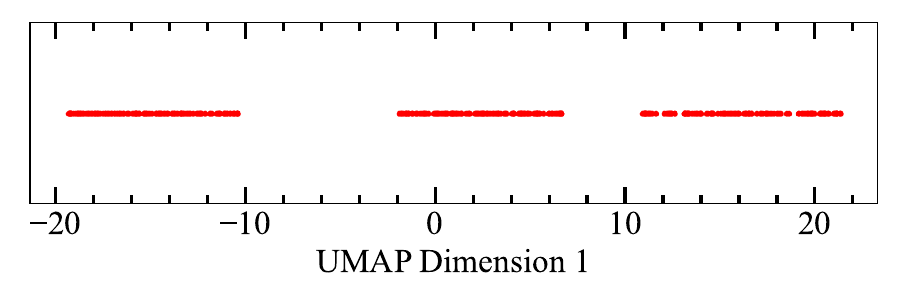}
        \end{minipage}
    }
    \caption{Visualization of dimensionality-reduction using UMAP when randomly generated data is projected from 2 dimensions to 1 dimension.}
    \label{Fig: Random data}
\end{figure}

\begin{figure}[htpb]
    \centering
    \includegraphics[scale=0.5]{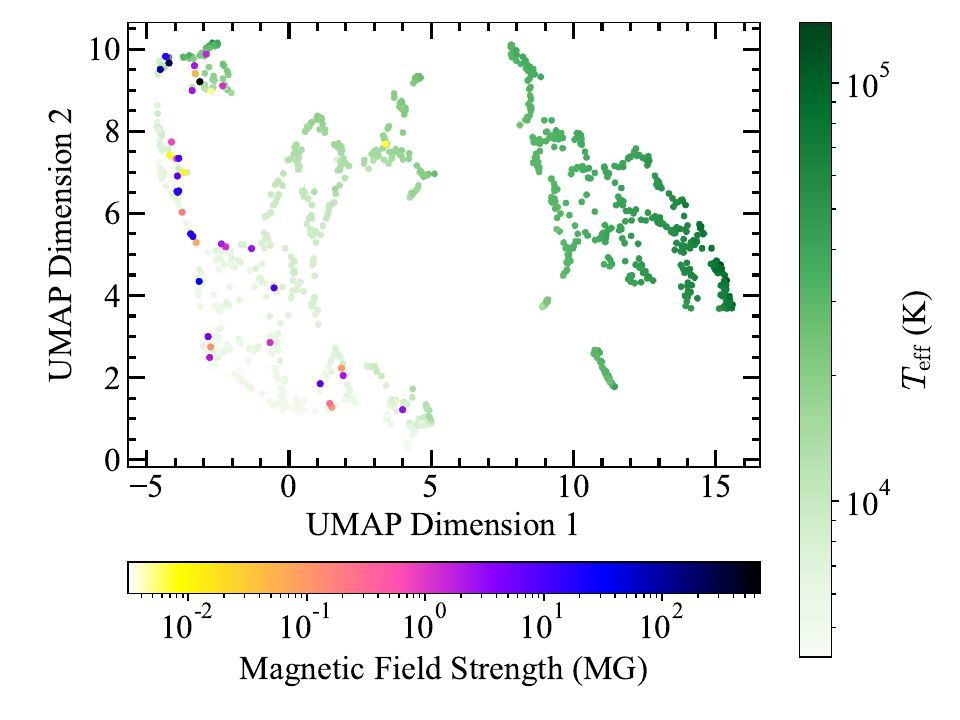}
    \caption{2-dimensional UMAP projection of the WD dataset constructed using five input parameters, excluding age, since age estimates largely depend on these parameters. The known MWDs remain confined to a single, well-defined cluster in the reduced parameter space.}
    \label{Fig: Test}
\end{figure}

It is worth emphasizing that the UMAP dimensionality-reduction technique is robust regardless of the presence or absence of correlation in the data. To exemplify this, we generate two sets of random 2-dimensional data: the first set exhibits high correlation, while the second shows lower correlation, as depicted in Fig.~\ref{Fig: Random data}. Naturally, both sets form three clusters in the original 2-dimensional space. Applying UMAP to reduce the data to 1 dimension, we observe that the clustering structure is preserved in both cases, demonstrating that UMAP effectively maintains the intrinsic data structure irrespective of correlation in data. Thus, adding or removing a parameter that is dependent on others does not qualitatively affect the overall outcome. For example, removing age as an intrinsic parameter does not change the conclusions: the known magnetized WDs still continue to occupy a single cluster, as depicted in Fig.~\ref{Fig: Test}. Nevertheless, we keep all six parameters as inputs because their inclusion allows a more uniform treatment with the full sample of all possible intrinsic parameters of WDs reported from direct or indirect measurements.

\subsection{Dimensionality Reduction with PCA}

\begin{figure}[htpb]
    \centering
    \includegraphics[scale=0.5]{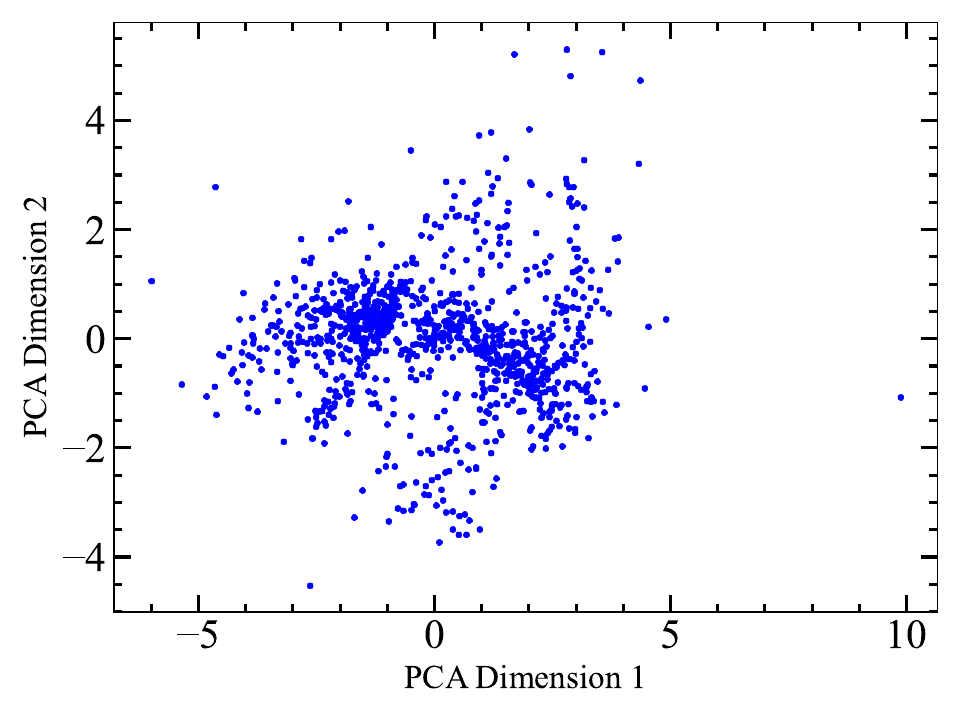}
    \caption{PCA dimensionality reduction.}
    \label{Fig: PCA}
\end{figure}

Before using UMAP, we tested PCA as an initial dimensionality reduction step. PCA captures linear correlations efficiently, but the WD parameter space is strongly nonlinear due to degeneracies among different parameters. Hence, PCA projections do not show any well-separated significant clusters, as evident from Fig.~\ref{Fig: PCA}, and we choose UMAP for our analysis.

\subsection{Clustering with DBSCAN}

In order to identify astrophysically distinct sub-populations from the UMAP-projected space, we use the DBSCAN algorithm that identifies clusters as adjacent regions of high density separated by areas of lower density. It requires two critical parameters. The first is \texttt{eps}, which represents the maximum distance between two samples for them to be considered as neighbors. We adopt \texttt{eps}=0.8 to ensure that different astrophysical groups are separated without excessive fragmentation. The second parameter is \texttt{min\_samples}, which represents the minimum number of points required to form a core cluster. We set \texttt{min\_samples}=4 to maintain the sensitivity for less populated clusters. As DBSCAN does not require the number of clusters to be specified in advance, it is very convenient for datasets in which the number of distinct populations is previously unknown. Note that DBSCAN is applied to the UMAP embedding rather than the original parameter space because DBSCAN relies on distance metrics that might become unreliable in high-dimensional spaces, which we cannot visualize. In the original feature space, distances between points might become increasingly uniform, reducing clustering contrast. Thus, it is not possible to visualize the clusters if we apply DBSCAN in the original high-dimensional space. Running DBSCAN on this embedding allows density-based clustering to operate on a space where meaningful distances are preserved.

\subsection{Magnetic field estimation with kNN regression}

\begin{figure}[htpb]
    \centering
    \includegraphics[scale=0.5]{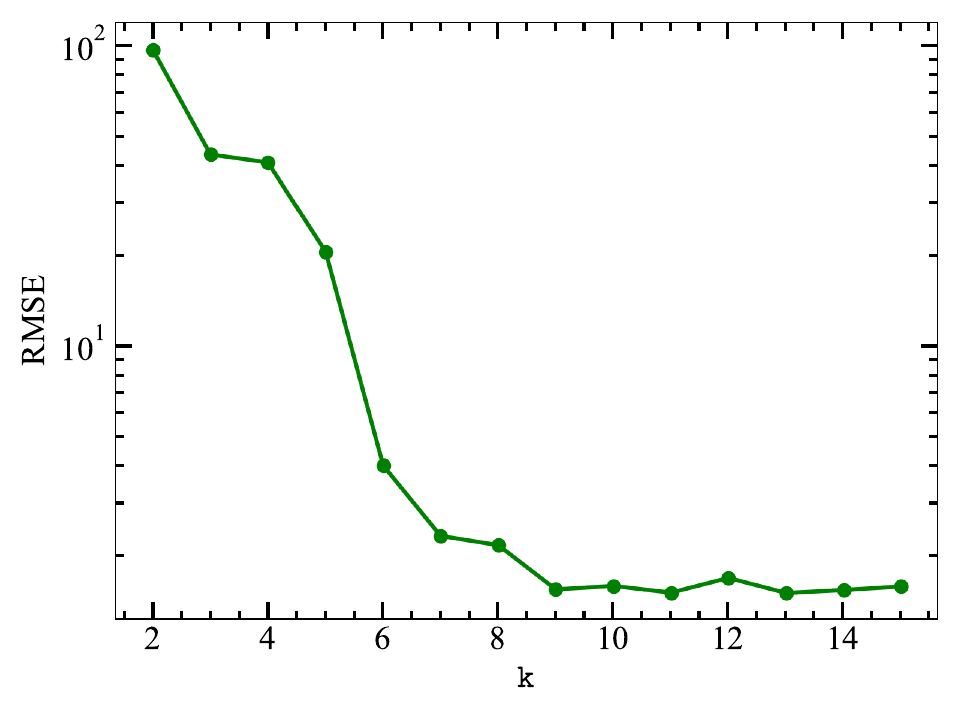}
    \caption{Change in RMSE with respect to \texttt{k} for a 90\%-10\% data split.}
    \label{Fig: Holdout}
\end{figure}

For WDs without magnetic field measurements, kNN regression is performed in the UMAP-optimized space. The assumption behind this method is that astrophysically similar objects are close in the vicinity. To test its reliability, we consider using a holdout validation set. As we have a relatively smaller number of known MWDs, the dataset is split into training (90\%) and test (10\%) subsets randomly. To reduce potential sampling bias associated with a single random split, it is repeated 1000 times for each \texttt{k}. The value of \texttt{k} represents that the code identifies \texttt{k} nearest neighbors with known magnetic fields. Thereby, we compute the root mean square error (RMSE), and the final performance estimate is obtained by averaging RMSE values across all repetitions, which is shown in Fig.~\ref{Fig: Holdout}.

We notice that around \texttt{k}=10, the error starts saturating, and thus we keep this number for computing magnetic fields of MWDs lacking field measurements. The estimated field strength is computed by taking an inverse-distance weighted average of all these values, given by
\begin{align}
    B_\mathrm{est} = \frac{\sum_i w_i B_i}{\sum_i w_i},
\end{align}
where $w_i = 1/(d_i+\epsilon)$ with $d_i$ is the Euclidean distance in UMAP space, $\epsilon$ is a small constant to avoid division by zero, and $B_i$ is the field strength of nearest neighbors with known measurements. Thus the associated uncertainty is given by
\begin{align}
\sigma_B = {\frac{\sqrt{\sum_i w_i^2\sigma_{B_i}^2}}{\sum_i w_i}},
\end{align}
where $\sigma_{B_i}$ is the uncertainty of known magnetic fields of each MWD. As kNN assumes local continuity of magnetic field properties and does not assume any parametric model for classification, its performance depends significantly on the density and representativeness of the training set.

\end{document}